\documentclass[final,3p,times]{elsarticle}

\usepackage{amsmath}
\usepackage{amssymb}
\usepackage{lipsum}
\usepackage{multicol}
\usepackage{fancyhdr}
\usepackage{xcolor} 

\definecolor{headercolor}{RGB}{128, 141, 180} 
\begin{document}

\begin{frontmatter}



\title{  Scattering approach for calculating one-loop effective action and vacuum energy}






 \author[label1]{Yuan-Yuan Liu}
 \affiliation[label1]{organization={Theoretical Physics Division, Chern Institute of Mathematics, Nankai University},
            city={Tianjin},
            postcode={300071},
            country={PR China}}

 \author[label2,label4]{Shi-Lin Li}
 \affiliation[label2]{organization={College of Physical Science and Technology, Bohai University},
            city={Jinzhou},
            postcode={121013},
            country={PR China}}

\author[label1,label4]{Yu-Jie Chen}
            
\author[label3]{Wen-Du Li} \ead{liwendu@tjnu.edu.cn}
 \affiliation[label3]{organization={College of Physics and Materials Science, Tianjin Normal University},
            city={Tianjin},
            postcode={300387},
            country={PR China}}

 \author[label4]{Wu-Sheng Dai}
 \ead{daiwusheng@tju.edu.cn}
 \affiliation[label4]{organization={Department of Physics, Tianjin University},
            city={Tianjin},
            postcode={300350},
            country={PR China}}

\begin{abstract}
We propose an approach for calculating one-loop effective actions and vacuum
energies in quantum field theory. Spectral functions are functions defined by
the eigenvalues of an operator. One-loop effective actions and vacuum energies
in quantum field theory, as well as scattering phase shifts and scattering
amplitudes in quantum mechanics, are all spectral functions. If a
transformation between different spectral functions is identified, we can
obtain a spectral function from another through the transformation. In this
paper, we convert quantum mechanical methods for calculating scattering phase
shifts and scattering amplitudes into quantum field theory methods for
calculating one-loop effective actions and vacuum energies.
\end{abstract}







\end{frontmatter}




\thispagestyle{fancy} 

\section{Introduction}

Physical information is embedded within physical operators, with classical
mechanics and classical field theory extracting classical information from
Hamiltonians, while quantum mechanics and quantum field theory extract quantum
information. For instance, the same Hamiltonian can be used to construct both
the Hamiltonian equation in classical mechanics and the Schr\"{o}dinger
equation in quantum mechanics, with the difference between them being the
method of extracting information.

The eigenproblem of a Hamiltonian,%

\begin{equation}
H\phi_{n}=\lambda_{n}\phi_{n},
\end{equation}
contains all of its information. If the eigenfunction $\phi_{n}$ and
corresponding eigenvalue $\lambda_{n}$ are known, the operator $H$ can be
fully determined. In the spectral representation, the operator can be
reconstructed by its eigenvalues and eigenfunctions: $H\left(  x^{\prime
},x\right)  =\sum_{n}\lambda_{n}\phi_{n}^{\ast}\left(  x^{\prime}\right)
\phi_{n}\left(  x\right)  $. Therefore, the eigenvalues and eigenfunctions
contain all of the operator's information.

The interesting question arises as to what information can be obtained solely
from knowledge of the eigenvalues without knowledge of the corresponding
eigenfunctions. This problem can be reformulated as follows: from the
eigenvalues of a Hamiltonian, what physical quantities can be constructed, and
what information can be extracted.

Spectral functions refer to physical quantities that can be constructed from
eigenvalues. The global heat kernel, also known as the partition function,
$K\left(  t\right)  =\sum_{n}\mathrm{e}^{-\lambda_{n}t}$, is an example of a
spectral function that is defined by the eigenvalue spectrum $\left\{
\lambda_{n}\right\}  $. Other significant spectral functions include
scattering phase shifts, one-loop effective actions, and vacuum energies.

The famous mathematical problem, "Can one hear the shape of a drum?"
originally posed by Kac \cite{kac1966can}, asks how much information can be
extracted from an eigenvalue spectrum. Essentially, Kac's question inquires
whether the Hamiltonian can be reconstructed solely from its eigenvalues. The
answer, however, is no \cite{gordon1992isospectral}. (Note that in Kac's
original problem, the information of the Hamiltonian is reflected in the
boundary condition, specifically the shape of the drum.)

\pagestyle{empty}

Since the eigenvalues alone do not provide all of the information of an
operator, the problem shifts towards determining "what information can be
extracted from eigenvalues." Various spectral functions can be defined using
eigenvalues, and different spectral functions can be transformed into one
another. Presently, the known transformations between spectral functions
include transformation among global heat kernels (partition functions),
one-loop effective actions, vacuum energies \cite{vassilevich2003heat}, the
transformation between global heat kernels and spectral counting functions
\cite{dai2009number,zhou2018calculating}, and transformation between
scattering phase shifts and global heat kernels
\cite{pang2012relation,li2015heat}. In this paper, we present the
transformations between scattering phase shifts and amplitudes, one-loop
effective actions, and vacuum energies.

The one-loop effective action and vacuum energy are quantities that arise in
quantum field theory, whereas the scattering phase shift and scattering
amplitude are quantities in quantum mechanics. All of these are considered
spectral functions. In this paper, we present a methodology for calculating
the one-loop effective action and vacuum energy in non-relativistic quantum
field theory using the scattering phase shift and scattering amplitude. This
approach enables the conversion of a quantum field theory problem into a
quantum mechanical problem or, equivalently, the translation of a quantum
mechanical method into a quantum field theory method. This methodology can be
extended to a relativistic theory by substituting the non-relativistic
scattering theory with a relativistic scattering theory.

Concretely, in Refs. \cite{dai2009number,zhou2018calculating}, we found the
relation between the global heat kernel and the spectral counting function
that counts the eigenstates whose eigenvalues are less than a certain number.
In Ref. \cite{pang2012relation}, we found the relation between the scattering
phase shift and the partial-wave global heat kernel, and in Ref.
\cite{li2015heat}, we found the relation between the scattering phase
shift\ and the local heat kernel, when we discussed the relationship between
the heat kernel method \cite{vassilevich2003heat} and the spectral method
\cite{weigel2018spectral} in quantum field theory. As an application, Ref.
\cite{liu2022seeley} proposes a method for calculating the scattering phase
shift based on the Seeley-DeWitt expansion in heat kernel theory. Based on
these previous works, in this paper, we give the relation between the
scattering phase shift and amplitude and the one-loop effective action and the
vacuum energy.

The relation between scattering phase shifts and amplitudes and one-loop
effective actions and vacuum energies converts quantum mechanical methods into
quantum-field-theory methods. By this relation, various methods for scattering
in quantum mechanics can be converted into methods for calculating one-loop
effective actions and vacuum energies in quantum field theory. As examples, in
the following, we convert two quantum mechanical methods, the Born
approximation and the WKB approximation, into methods for calculating one-loop
effective actions and vacuum energies.

The heat kernel serves as a bridge in our method, which bridges quantum field
theory and quantum mechanics. The heat kernel expansion is an important method
in quantum field theory. There are two heat kernel expansions: the covariant
perturbation theory
\cite{barvinsky1987beyond,barvinsky1990covariant,barvinsky1990covariant3,avramidi1990nonlocal,gusev2009heat,avramidi2015heat,mukhanov2007introduction}
and the Schwinger-DeWitt technique
\cite{barvinsky1985generalized,vassilevich2003heat}. Various methods are
developed for heat kernel approaches, such as calculating heat kernel traces
by the path integral \cite{bastianelli2020path}, the Green function approach
\cite{yannick2020heat}, the technique of labeled operators
\cite{salcedo2004covariant}, and heat kernel diagrammatic equations
\cite{codello2013non}. The heat kernel of higher-order differential operators
is considered \cite{barvinsky2019heat}. Heat kernel expansions are very
important, such as the Schwinger-DeWitt expansion in the induced gravity on
the AdS background \cite{altshuler2015sakharov}. The heat kernel method
applies to calculate effective actions \cite{avramidi2002heat}, such as the
effective field theory in curved spacetime
\cite{ruf2018quantum,nakonieczny2019curved}, the heat kernel expansion and the
one-loop effective action in QCD \cite{megias2004thermal}, the Seeley-DeWitt
expansion for the one-loop effective action in the Einstein-Maxwell theory
\cite{karan2019seeley}, the one-loop effective action for the modified
Gauss-Bonnet gravity \cite{cognola2009one} and in dS$_{2}$ and AdS$_{2}$
spacetime \cite{cai2014one}, $\varphi^{4}$-fields \cite{vassilevich2005heat}%
,\ and various operators
\cite{drozd2016universal,kramer2020completing,fucci2012heat}. The heat kernel
method also applies to calculate vacuum energies, such as Casimir energies in
curved spacetime
\cite{sorge2014casimir,assel2015casimir,zhang2015thermal,zhang2017theoretical}
and in spherically symmetric backgrounds \cite{beauregard2015casimir}. The
vacuum energy is also calculated by the spectral functions
\cite{fulling2007vacuum,fulling2003systematics}. Applying scattering theory to
calculate vacuum energy is pioneered in Refs.
\cite{graham1999energy,graham2002calculating,graham2003casimir,graham2003negative,rahi2009scattering,weigel2018spectral}%
. Various methods in scattering theory can be found in Refs.
\cite{newton2013scattering,friedrich2013scattering,lax2016scattering,belkic2020principles}%
. Reviews on one-loop effective action can be found in Refs.
\cite{buchbinder1992effective,bytsenko1994zeta,nojiri2001quantum,buchbinder1989effective,odintsov1990parametrization}%
. The one-loop calculation on different backgrounds
\cite{bamba2014one,bytsenko1995large,brevik2001quantum} and the two-loop
effective action in quantum gravity \cite{buchbinder1992two} are considered.

In section \ref{briefreview}, we give a brief review of various spectral
functions. In section \ref{HOVPS}, we calculate global heat kernels, one-loop
effective actions, and vacuum energies from scattering phase shifts. In
sections \ref{BA3D} and \ref{BAnD}, we convert the Born approximation method
into a method for calculating one-loop effective actions and vacuum energies
in three and $n$ dimensions, respectively. In section \ref{WKB}, we convert
the WKB approximation method into a method for calculating one-loop effective
actions and vacuum energies. In section \ref{HOVPA}, we calculate global heat
kernels, one-loop effective actions, and vacuum energies from scattering
amplitudes. The conclusion is given in section \ref{Conclusion}. In Appendix
\ref{integralrepresentation}, we give some integral representations for the
Bessel function.

\section{Heat kernel, one-loop effective action, vacuum energy, and scattering
phase shift: brief review \label{briefreview}}

In this section, we briefly introduce the scattering phase shift, the heat
kernel, the one-loop effective action, and the vacuum energy. They are
spectral functions determined by the eigenvalue.

\subsection{Scattering phase shift}

We consider the elastic scattering of a plane wave on a spherically symmetric potential.

If there is no scattering, namely $V\left(  r\right)  =0$, the wave function
is a plane wave \cite{li2016scattering,liu2014scattering}:%
\begin{equation}
\psi_{0}\left(  r,\theta\right)  =\mathrm{e}^{\mathrm{i}kr\cos\theta}%
=\sum_{l=0}^{\infty}\left(  2l+1\right)  \mathrm{i}^{l}\frac{1}{2}\left[
h_{l}^{\left(  2\right)  }\left(  kr\right)  +h_{l}^{\left(  1\right)
}\left(  kr\right)  \right]  P_{l}\left(  \cos\theta\right)  , \label{psi0}%
\end{equation}
where $k$ is the incident momentum, $\theta$ is the scattering angle. For
scattering, we need to impose scattering boundary condition. For short-range
potentials, the scattering boundary condition is\cite{friedrich2013scattering}%
\[
\psi\left(  r,\theta\right)  =\mathrm{e}^{\mathrm{i}kr\cos\theta}+f\left(
\theta\right)  \frac{\mathrm{e}^{\mathrm{i}kr}}{r}.
\]
Expanding the wave function into partial waves gives
\cite{li2016scattering,liu2014scattering}
\begin{equation}
\psi\left(  r,\theta\right)  =\sum_{l=0}^{\infty}\left(  2l+1\right)
\mathrm{i}^{l}\frac{1}{2}\left[  h_{l}^{\left(  2\right)  }\left(  kr\right)
+\mathrm{e}^{2\mathrm{i}\delta_{l}}h_{l}^{\left(  1\right)  }\left(
kr\right)  \right]  P_{l}\left(  \cos\theta\right)  , \label{psi}%
\end{equation}
where $h_{l}^{\left(  1\right)  }\left(  kr\right)  $, the first kind
spherical Hankel function, describes the outgoing wave, and $h_{l}^{\left(
2\right)  }\left(  kr\right)  $, the second kind spherical Hankel function,
describes the incoming wave. Comparing $\psi_{0}\left(  r,\theta\right)  $ and
$\psi\left(  r,\theta\right)  $, we can see that the effect of scattering is
to multiply the outgoing wave by a phase factor $\mathrm{e}^{2\mathrm{i}%
\delta_{l}}$. Scattering causes a phase shift $\delta_{l}$ on the outgoing
wave function, called the scattering phase shift.

If the observer is far from the target, one can take the large-distance
asymptotic approximation \cite{li2016scattering}:%
\begin{equation}
h_{l}^{\left(  1,2\right)  }\left(  kr\right)  \overset{r\rightarrow
\infty}{\sim}\left(  \mp\mathrm{i}\right)  ^{l+1}\frac{\mathrm{e}%
^{\pm\mathrm{i}kr}}{kr}. \label{h12inf}%
\end{equation}

Under the large-distance asymptotic approximation, the incident plane wave
(\ref{psi0}) becomes%
\begin{equation}
\psi_{0}\left(  r,\theta\right)  \overset{r\rightarrow\infty}{\sim}\sum
_{l=0}^{\infty}\left(  2l+1\right)  \mathrm{i}^{l}\frac{\sin\left(
kr-l\pi/2\right)  }{kr}P_{l}\left(  \cos\theta\right)  ,
\end{equation}
and the scattering wave (\ref{psi}) becomes%
\begin{equation}
\psi\left(  r,\theta\right)  \overset{r\rightarrow\infty}{\sim}\sum
_{l=0}^{\infty}\left(  2l+1\right)  \mathrm{i}^{l}\mathrm{e}^{\mathrm{i}%
\delta_{l}}\frac{\sin\left(  kr-l\pi/2+\delta_{l}\right)  }{kr}P_{l}\left(
\cos\theta\right)  .
\end{equation}

The corresponding radial wave functions, under the large-distance asymptotic
approximation, are%
\begin{equation}
R_{l}^{0}\left(  r\right)  \overset{r\rightarrow\infty}{\sim}\frac{\sin\left(
kr-l\pi/2\right)  }{kr},
\end{equation}
and%
\begin{equation}
R_{l}\left(  r\right)  \overset{r\rightarrow\infty}{\sim}\frac{\sin\left(
kr-l\pi/2+\delta_{l}\left(  k\right)  \right)  }{kr}.
\end{equation}

\subsection{Global Heat kernel, one-loop effective action, and vacuum energy}

For operator $D$, the spectral function is defined by its eigenvalues
$\left\{  \lambda_{n}\right\}  $. Formally, the global heat kernel is
\cite{vassilevich2003heat}%
\begin{equation}
K\left(  t\right)  =\sum_{n}e^{-\lambda_{n}t}, \label{Kt}%
\end{equation}
the one-loop effective action is \cite{vassilevich2003heat,dai2010approach}%
\begin{equation}
W=\sum_{n}\ln\sqrt{\lambda_{n}}, \label{W}%
\end{equation}
and the vacuum energy is \cite{elizalde2012ten}%
\begin{equation}
E_{0}=\frac{1}{2}\sum_{n}\lambda_{n}. \label{E0}%
\end{equation}

Physical operators, e.g., the Hamiltonian, are lower-bounded. The global heat
kernel (\ref{Kt})\ is well-defined. Nevertheless, the one-loop effective
action (\ref{W}) and the vacuum energy (\ref{E0})\ are not well-defined, for
they diverge for upper unbounded spectra. In order to obtain finite one-loop
effective actions and vacuum energies, one introduces the regularized one-loop
effective action and the regularized vacuum energy \cite{elizalde2012ten}.

By inspecting the formal expressions of global heat kernels, one-loop
effective actions, and\ vacuum energies, Eqs. (\ref{Kt}), (\ref{W}), and
(\ref{E0}), we can see that they are related by the relations $W=-\frac{1}%
{2}\int_{0}^{\infty}\frac{1}{t}K\left(  t\right)  \mathrm{d}t$\ and
$E_{0}=\frac{1}{2}\frac{1}{\Gamma\left(  -1\right)  }\int_{0}^{\infty}K\left(
t\right)  t^{-2}\mathrm{d}t$. However, it is obvious that these two relations
also diverge. To remove divergence, by using the well-defined global heat
kernel, one introduces the regularized one-loop effective action
\cite{vassilevich2003heat,dai2010approach}%
\begin{equation}
W\left(  s\right)  =-\frac{1}{2}\tilde{\mu}^{2s}\int_{0}^{\infty}K\left(
t\right)  t^{s-1}\mathrm{d}t \label{wsandrehe}%
\end{equation}
and the regularized vacuum energy%
\begin{equation}
E_{0}\left(  \epsilon\right)  =\frac{1}{2}\tilde{\mu}^{2\epsilon}\frac
{1}{\Gamma\left(  -\frac{1}{2}+\epsilon\right)  }\int_{0}^{\infty}K\left(
t\right)  t^{-\frac{1}{2}+\epsilon-1}\mathrm{d}t, \label{eandrehe}%
\end{equation}
where $\tilde{\mu}$ is a constant of the dimension of mass introduced to keep
the proper dimension \cite{vassilevich2003heat}. When $s=0$ and $\epsilon
=-1/2$, the regularized one-loop effective action and vacuum energy, $W\left(
s\right)  $ and $E_{0}\left(  \epsilon\right)  $, recover one-loop effective
action and vacuum energy, $W$ and $E_{0}$, but, of course, such a substitution
must undergo a regularization process. Moreover, Ref. \cite{blau1988zeta}
considers the vacuum energy in static and unbounded spacetime and provides a
finite definition of vacuum energy using the minimal subtraction scheme.
Various techniques for subtracting divergences can also be found in Ref.
\cite{zhang2022renormalization}.

In this paper, we suggest an approach for calculating the heat kernel, the
regularized one-loop effective action, and the regularized vacuum energy from
the scattering phase shift and the scattering amplitude.

\section{Calculating global heat kernel, one-loop effective action, and vacuum
energy from scattering phase shift \label{HOVPS}}

In the section, we calculate the global heat kernel, the regularized one-loop
effective action, and the regularized vacuum energy from the scattering phase shift.

The local heat kernel $K\left(  t;\mathbf{r},\mathbf{r}^{\prime}\right)  $ is
the Green function of the initial-value problem for the operator
$D=-\nabla^{2}+V\left(  r\right)  $. The local partial-wave heat kernel
$K_{l}\left(  t;r,r^{\prime}\right)  $ is the Green function of the
initial-value problem for the radial operator $D_{l}=-\frac{1}{r^{2}}%
\frac{\mathrm{d}}{\mathrm{d}r}\left(  r^{2}\frac{\mathrm{d}}{\mathrm{d}%
r}\right)  +\frac{l\left(  l+1\right)  }{r^{2}}+V\left(  r\right)  $. The
global heat kernel $K\left(  t\right)  $\ is the trace of $K\left(
t;\mathbf{r},\mathbf{r}^{\prime}\right)  $, and the global partial-wave heat
kernel $K_{l}\left(  t\right)  $ is the trace of $K_{l}\left(  t;r,r^{\prime
}\right)  $.

The global partial-wave heat kernel can be divided into three parts:
$K_{l}\left(  t\right)  =K_{l}^{\text{s}}\left(  t\right)  +K_{l}%
^{\text{free}}\left(  t\right)  +K_{l}^{\text{bound}}\left(  t\right)  $,
where $K_{l}^{\text{s}}\left(  t\right)  $ is the scattering-state global
partial-wave heat kernel, i.e.
\begin{equation}
K_{l}^{\text{s}}\left(  t\right)  =\int_{0}^{\infty}\rho\left(  \lambda
\right)  e^{-\lambda t}\mathrm{d}t, \label{hkscat}%
\end{equation}
where $\rho\left(  \lambda\right)  $ is the state density. The summation over
the continuum spectrum of scattering states becomes an integral.
$K_{l}^{\text{free}}\left(  t\right)  =\left(  4\pi t\right)  ^{-3/2}$ is the
free global partial-wave heat kernel, and $K_{l}^{\text{bound}}\left(
t\right)  $ is the bound-state global partial-wave heat kernel, i.e.
\begin{equation}
K_{l}^{\text{bound}}\left(  t\right)  =\sum_{\text{all bound states}%
}e^{-\lambda_{n}t},
\end{equation}
where the summation runs over all bound states. The three-dimensional free
global partial-wave heat kernel $K_{l}^{\text{free}}\left(  t\right)
=R/\sqrt{4\pi t}-\left(  l+1/2\right)  /2$. The bound-state global
partial-wave heat kernel $K_{l}^{\text{bound}}\left(  t\right)  $ needs to
work out the sum over all discrete bound states. In this paper, we care only
about the scattering states, for most quantum-field-theory problems are
related to scattering states. In Ref. \cite{pang2012relation}, as a by-product
of the discussion of the relation between the heat kernel method and
the\ spectral method in quantum field theory, we find a relation between the
partial-wave phase shift and the global partial-wave heat kernel of scattering
states:%
\begin{equation}
K_{l}^{\text{s}}\left(  t\right)  =\frac{2}{\pi}t\int_{0}^{\infty}\delta
_{l}\left(  k\right)  \mathrm{e}^{-k^{2}t}k\mathrm{d}k-\frac{\delta_{l}\left(
0\right)  }{\pi}, \label{hkbyps}%
\end{equation}
where the superscript $s$ denotes the scattering-state heat kernel. According
to the Levinson theorem \cite{ma2006levinson,kellendonk2008topological},
$\delta_{l}\left(  0\right)  =n_{l}\pi$, where $n_{l}$ the number of bound
states with the angular momentum $l$. If there exists the half-bound state
(the half-bound state may occur only when $l=0$), $\delta_{l}\left(  0\right)
=\left(  n_{l}+\frac{1}{2}\right)  \pi$. If only considering scattering
states, $\delta_{l}\left(  0\right)  $ does not contribute.

The relation between the global heat kernel $K^{\text{s}}\left(  t\right)  $
and the global partial-wave heat kernel $K_{l}^{\text{s}}\left(  t\right)  $
of scattering states is%
\begin{equation}
K^{\text{s}}\left(  t\right)  =\sum_{l=0}^{\infty}D_{l}K_{l}^{\text{s}}\left(
t\right)  , \label{Kst}%
\end{equation}
where $D_{l}$ is the degeneracy. Then, by Eqs. (\ref{hkbyps}) and (\ref{Kst}),
we can obtain the relation between the global heat kernel and the scattering
phase shift:%
\begin{equation}
K^{\text{s}}\left(  t\right)  =\frac{t}{\pi}\sum_{l=0}^{\infty}D_{l}\int%
_{0}^{\infty}\mathrm{d}k^{2}\mathrm{e}^{-k^{2}t}\delta_{l}\left(  k\right)  .
\label{psdaorehe}%
\end{equation}

The relation between the regularized one-loop effective action and the global
heat kernel can be obtained by Eqs. (\ref{wsandrehe}) and (\ref{psdaorehe}):
\begin{align}
W^{\text{s}}\left(  s\right)   &  =-\frac{1}{2\pi}\tilde{\mu}^{2s}\int%
_{0}^{\infty}t^{s}\mathrm{d}t\int_{0}^{\infty}\mathrm{d}k^{2}\mathrm{e}%
^{-k^{2}t}\sum_{l=0}^{\infty}D_{l}\delta_{l}\left(  k\right) \nonumber\\
&  =-\frac{1}{2\pi}\tilde{\mu}^{2s}\Gamma\left(  s+1\right)  \sum
_{l=0}^{\infty}D_{l}\int_{0}^{\infty}\mathrm{d}k^{2}\frac{\delta_{l}\left(
k\right)  }{\left(  k^{2}\right)  ^{s+1}}. \label{1loop}%
\end{align}

The relation between the regularized vacuum energy and the global heat kernel
can be obtained by Eqs. (\ref{eandrehe}) and (\ref{psdaorehe}):%
\begin{align}
E_{0}^{\text{s}}\left(  \epsilon\right)   &  =\frac{1}{2\pi}\tilde{\mu
}^{2\epsilon}\frac{1}{\Gamma\left(  -\frac{1}{2}+\epsilon\right)  }\int%
_{0}^{\infty}t^{-\frac{1}{2}+\epsilon}\mathrm{d}t\int_{0}^{\infty}%
\mathrm{d}k^{2}\mathrm{e}^{-k^{2}t}\sum_{l=0}^{\infty}D_{l}\delta_{l}\left(
k\right) \nonumber\\
&  =\frac{1}{2\pi}\tilde{\mu}^{2\epsilon}\frac{\Gamma\left(  \frac{1}%
{2}+\epsilon\right)  }{\Gamma\left(  -\frac{1}{2}+\epsilon\right)  }\sum
_{l=0}^{\infty}D_{l}\int_{0}^{\infty}\mathrm{d}k^{2}\frac{\delta_{l}\left(
k\right)  }{\left(  k^{2}\right)  ^{1/2+\epsilon}}. \label{vE}%
\end{align}

The relations (\ref{1loop})\ and (\ref{vE}) convert the method for calculating
scattering phase shifts in quantum mechanics into a method for calculating
one-loop effective actions and vacuum energies in quantum field theory. That
is, it converts a quantum-field-theory problem into a quantum-mechanical problem.

In Eqs. (\ref{psdaorehe}), (\ref{1loop}),\ and (\ref{vE}), there is a sum
should be worked out. In order to deal with this sum, in Appendix
(\ref{integralrepresentation}) we give some integral representations for the
Bessel function.

Ref. \cite{weigel2018spectral} provides the relation between the scattering
phase shift and the density of states $\rho_{l}\left(  k\right)  $ and the
Jost function $F_{l}\left(  k\right)  $:
\begin{equation}
\frac{1}{\pi}\frac{\mathrm{d}\delta_{l}}{\mathrm{d}k}=\rho_{l}\left(
k\right)  -\rho_{l}^{\left(  0\right)  }\left(  k\right)
\end{equation}
and%
\begin{equation}
e^{2i\delta_{l}}=\frac{F_{l}\left(  -k\right)  }{F_{l}\left(  k\right)  }.
\end{equation}
This allows us to calculate the density of states and the phase of the Jost
function from the scattering phase shift.

\section{Born approximation: three-dimensional case \label{BA3D}}

The integral equation method is a fundamental approach for scattering in
quantum mechanics and in quantum field theory. It can also be applied to the
scattering problem in curved spacetime \cite{li2018scalar,li2021scalar}. By
the Green function, the integral equation method converts the differential
equation defined by the operator $D$, e.g., the eigenequation of the
Hamiltonian, into an integral equation. This integral equation can be solved
as a perturbation series by the iterative method. The leading-order
contribution of the perturbation solution is the Born approximation. The Born
approximation is the most important and mature method in the perturbation
theory of scattering.

In the following, we convert the Born approximation of calculating scattering
phase shifts in quantum mechanics into a method of calculating one-loop
effective actions and vacuum energies in quantum field theory.

Due to the high accuracy of Born approximation at large $k$, for the
regularization of ultraviolet divergences (divergences at large $k$), we can
directly employ subtraction of the first- or second-order Born approximation
scattering phase shift to improve the convergence of the integral. Therefore,
considering the results of the first two orders of Born approximation is
meaningful for the regularization of ultraviolet divergences
\cite{weigel2018spectral,graham2002calculating}.

\subsection{First-order Born approximation}

The first-order Born approximation of the scattering phase shift for a
spherically symmetric potential $V\left(  r\right)  $ in three dimensions is
\cite{joachain1975quantum}%
\begin{equation}
\delta_{l}^{\left(  1\right)  }=-\frac{\pi}{2}\int_{0}^{\infty}J_{l+1/2}%
^{2}\left(  kr\right)  V\left(  r\right)  r\mathrm{d}r, \label{born1}%
\end{equation}
where $J_{\nu}\left(  z\right)  $ is the Bessel function.

\subsubsection{Heat kernel}

The first-order approximation of the heat kernel can be obtained by
substituting Eq. (\ref{born1}) and the degeneracy in three dimensions,
$D_{l}=2l+1$, into Eq. (\ref{psdaorehe}):%
\begin{equation}
K^{\text{s}\left(  1\right)  }\left(  t\right)  =-\frac{t}{2}\int_{0}^{\infty
}V\left(  r\right)  r\mathrm{d}r\int_{0}^{\infty}\mathrm{d}k^{2}%
\mathrm{e}^{-k^{2}t}\sum_{l=0}^{\infty}\left(  2l+1\right)  J_{l+1/2}%
^{2}\left(  kr\right)  .
\end{equation}
By the sum rule \cite{graham2009spectral}%
\begin{equation}
\sum_{l=0}^{\infty}\frac{\left(  2q+2l\right)  \Gamma\left(  2q+l\right)
}{\Gamma\left(  l+1\right)  }J_{q+l}^{2}\left(  z\right)  =\frac{\Gamma\left(
2q+1\right)  }{\Gamma\left(  q+1\right)  ^{2}}\left(  \frac{z}{2}\right)
^{2q},
\end{equation}
we arrive at%
\begin{equation}
\sum_{l=0}^{\infty}\left(  2l+1\right)  J_{l+1/2}^{2}\left(  kr\right)
=\frac{2}{\pi}kr. \label{sum1}%
\end{equation}
Thus, we have%
\begin{equation}
K^{\text{s}\left(  1\right)  }\left(  t\right)  =-\frac{t}{2}\int_{0}^{\infty
}V\left(  r\right)  r\mathrm{d}r\int_{0}^{\infty}\mathrm{d}k^{2}%
\mathrm{e}^{-k^{2}t}\frac{2kr}{\pi}.
\end{equation}
Working out the integral that is a Laplace transform gives the first-order
Born approximation for heat kernels:%
\begin{equation}
K^{\text{s}\left(  1\right)  }\left(  t\right)  =-\frac{1}{\sqrt{4\pi t}}%
\int_{0}^{\infty}V\left(  r\right)  r^{2}\mathrm{d}r. \label{ktborn1}%
\end{equation}

\subsubsection{One-loop effective action}

The first-order one-loop effective action can be obtained by substituting the
first-order phase shift (\ref{born1}) into Eq. (\ref{1loop}):%
\begin{equation}
W^{\left(  1\right)  }\left(  s\right)  =\frac{1}{4}\tilde{\mu}^{2s}%
\Gamma\left(  s+1\right)  \int_{0}^{\infty}V\left(  r\right)  r\mathrm{d}%
r\int_{0}^{\infty}\left(  k^{2}\right)  ^{-s-1}\left[  \sum_{l=0}^{\infty
}\left(  2l+1\right)  J_{l+1/2}^{2}\left(  kr\right)  \right]  \mathrm{d}%
k^{2}.
\end{equation}
Using the sum rule (\ref{sum1}), we arrive at%
\begin{equation}
W^{\left(  1\right)  }\left(  s\right)  =\frac{1}{2\pi}\tilde{\mu}^{2s}%
\Gamma\left(  s+1\right)  \int_{0}^{\infty}V\left(  r\right)  r^{2}%
\mathrm{d}r\int_{0}^{\infty}\left(  k^{2}\right)  ^{-s-1}k\mathrm{d}k^{2}.
\end{equation}
Here the integral of $k$ may diverge. According to Ref.
\cite{vassilevich2005heat}, we rewrite $\left(  k^{2}\right)  ^{-s-1}$ as
$\left(  k^{2}+m^{2}\right)  ^{-s-1}$:%
\begin{equation}
W^{\left(  1\right)  }\left(  s\right)  =\frac{1}{2\pi}\tilde{\mu}^{2s}%
\Gamma\left(  s+1\right)  \int_{0}^{\infty}V\left(  r\right)  r^{2}%
\mathrm{d}r\int_{0}^{\infty}\left(  k^{2}+m^{2}\right)  ^{-s-1}k\mathrm{d}%
k^{2}.
\end{equation}
Working out the integral, we obtain the first-order Born approximation of the
one-loop effective action:%
\begin{equation}
W^{\left(  1\right)  }\left(  s\right)  =\frac{\tilde{\mu}^{2s}\Gamma\left(
s-\frac{1}{2}\right)  }{4\sqrt{\pi}}\left(  m^{2}\right)  ^{\frac{1}{2}-s}%
\int_{0}^{\infty}V\left(  r\right)  r^{2}\mathrm{d}r.
\end{equation}

\subsubsection{Vacuum energy}

The first-order vacuum energy can be obtained by substituting the first-order
phase shift (\ref{born1}) into Eq. (\ref{vE}):%
\begin{equation}
E_{0}^{\left(  1\right)  }\left(  \epsilon\right)  =-\frac{1}{4}\tilde{\mu
}^{2\epsilon}\frac{\Gamma\left(  \frac{1}{2}+\epsilon\right)  }{\Gamma\left(
-\frac{1}{2}+\epsilon\right)  }\int_{0}^{\infty}V\left(  r\right)
r\mathrm{d}r\int_{0}^{\infty}\left(  k^{2}\right)  ^{-1/2-\epsilon}\left[
\sum_{l=0}^{\infty}\left(  2l+1\right)  J_{l+1/2}^{2}\left(  kr\right)
\right]  \mathrm{d}k^{2}.
\end{equation}
Using the sum rule (\ref{sum1}), we arrive at%
\begin{equation}
E_{0}^{\left(  1\right)  }\left(  \epsilon\right)  =-\frac{1}{2\pi}\tilde{\mu
}^{2\epsilon}\frac{\Gamma\left(  \frac{1}{2}+\epsilon\right)  }{\Gamma\left(
-\frac{1}{2}+\epsilon\right)  }\int_{0}^{\infty}V\left(  r\right)
r^{2}\mathrm{d}r\int_{0}^{\infty}\left(  k^{2}\right)  ^{-1/2-\epsilon
}k\mathrm{d}k^{2}.
\end{equation}
The integral of $k$ may diverge. According to Ref. \cite{vassilevich2005heat},
we rewrite $\left(  k^{2}\right)  ^{-1/2-\epsilon}$ as $\left(  k^{2}%
+m^{2}\right)  ^{-1/2-\epsilon}$:%
\begin{equation}
E_{0}^{\left(  1\right)  }\left(  \epsilon\right)  =-\frac{1}{2\pi}\tilde{\mu
}^{2\epsilon}\frac{\Gamma\left(  \frac{1}{2}+\epsilon\right)  }{\Gamma\left(
-\frac{1}{2}+\epsilon\right)  }\int_{0}^{\infty}V\left(  r\right)
r^{2}\mathrm{d}r\int_{0}^{\infty}\left(  k^{2}+m^{2}\right)  ^{-1/2-\epsilon
}k\mathrm{d}k^{2}.
\end{equation}
Working out the integral, we obtain the first-order Born approximation of the
vacuum energy:%
\begin{equation}
E_{0}^{\left(  1\right)  }\left(  \epsilon\right)  =-\frac{\tilde{\mu
}^{2\epsilon}}{4\sqrt{\pi}}\frac{\Gamma\left(  \epsilon-1\right)  }%
{\Gamma\left(  -\frac{1}{2}+\epsilon\right)  }\left(  m^{2}\right)
^{1-\epsilon}\int_{0}^{\infty}V\left(  r\right)  r^{2}\mathrm{d}r.
\end{equation}

\subsection{Second-order Born approximation}

The second-order Born approximation of the scattering phase shift for a
spherically symmetric potential $V\left(  r\right)  $ is
\cite{joachain1975quantum}%
\begin{align}
\delta_{l}^{\left(  2\right)  }  &  =-\int_{0}^{\infty}kj_{l}\left(
kr\right)  n_{l}\left(  kr\right)  V\left(  r\right)  r^{2}\mathrm{d}r\int%
_{0}^{r}kj_{l}^{2}\left(  kr^{\prime}\right)  V\left(  r^{\prime}\right)
r^{\prime2}\mathrm{d}r^{\prime}\nonumber\\
&  -\int_{0}^{\infty}kj_{l}^{2}\left(  kr\right)  V\left(  r\right)
r^{2}\mathrm{d}r\int_{r}^{\infty}kj_{l}\left(  kr^{\prime}\right)
n_{l}\left(  kr^{\prime}\right)  V\left(  r^{\prime}\right)  r^{\prime
2}\mathrm{d}r^{\prime}, \label{born2}%
\end{align}
where $j_{\nu}\left(  z\right)  $ and $n_{\nu}\left(  z\right)  $ are the
spherical Bessel functions of the first kind and of the second kind, respectively.

\subsubsection{Heat kernel}

The second-order approximation of the heat kernel can be obtained by
substituting the second-order phase shift\ (\ref{born2}) into Eq.
(\ref{psdaorehe}):%
\begin{align}
K^{\text{s}\left(  2\right)  }\left(  t\right)   &  =-\frac{t}{\pi}\int%
_{0}^{\infty}\mathrm{e}^{-k^{2}t}k^{2}\mathrm{d}k^{2}\int_{0}^{\infty}V\left(
r\right)  r^{2}\mathrm{d}r\int_{0}^{r}\mathrm{d}r^{\prime}r^{\prime2}V\left(
r^{\prime}\right)  \Sigma_{1}\left(  k;r,r^{\prime}\right) \nonumber\\
&  -\frac{t}{\pi}\int_{0}^{\infty}\mathrm{e}^{-k^{2}t}k^{2}\mathrm{d}k^{2}%
\int_{0}^{\infty}V\left(  r\right)  r^{2}\mathrm{d}r\int_{r}^{\infty
}\mathrm{d}r^{\prime}r^{\prime2}V\left(  r^{\prime}\right)  \Sigma_{2}\left(
k;r,r^{\prime}\right)  , \label{kb2}%
\end{align}
where%
\begin{align}
\Sigma_{1}\left(  k;r,r^{\prime}\right)   &  =\sum_{l=0}^{\infty}\left(
2l+1\right)  j_{l}\left(  kr\right)  n_{l}\left(  kr\right)  j_{l}^{2}\left(
kr^{\prime}\right)  ,\label{W1krr}\\
\Sigma_{2}\left(  k;r,r^{\prime}\right)   &  =\sum_{l=0}^{\infty}\left(
2l+1\right)  j_{l}^{2}\left(  kr\right)  j_{l}\left(  kr^{\prime}\right)
n_{l}\left(  kr^{\prime}\right)  . \label{W2krr}%
\end{align}
In the following, we deal with the above sums.

To perform these sums, we give an integral representation of $j_{l}^{2}\left(
kr\right)  $ in Appendix \ref{Appendix1}:%
\begin{equation}
j_{l}^{2}\left(  kr\right)  =\frac{1}{2}\int_{0}^{\pi}\frac{\sin qr}{qr}%
P_{l}\left(  \cos\theta\right)  \mathrm{d}\cos\theta
\end{equation}
and an integral representation of $j_{l}\left(  kr\right)  n_{l}\left(
kr\right)  $ in Appendix \ref{Appendix2}:%
\begin{equation}
j_{l}\left(  kr\right)  n_{l}\left(  kr\right)  =-\frac{1}{2}\int_{0}^{\pi
}\frac{\cos qr}{qr}P_{l}\left(  \cos\theta\right)  \mathrm{d}\cos\theta.
\end{equation}

Substituting the above two integral representations into Eq. (\ref{W1krr})
gives
\begin{equation}
\Sigma_{1}\left(  k;r,r^{\prime}\right)  =-\frac{1}{4}\int_{0}^{\pi}\frac{\cos
qr}{qr}\mathrm{d}\cos\theta\int_{0}^{\pi}\mathrm{d}\cos\theta^{\prime}%
\frac{\sin q^{\prime}r^{\prime}}{q^{\prime}r^{\prime}}\sum_{l=0}^{\infty
}\left(  2l+1\right)  P_{l}\left(  \cos\theta\right)  P_{l}\left(  \cos
\theta^{\prime}\right)  , \label{w11}%
\end{equation}
where $q=2k\sin\frac{\theta}{2}$ and $q^{\prime}=2k\sin\frac{\theta^{\prime}%
}{2}$. Using the relation \cite{olver2010nist}%
\begin{equation}
\sum_{l=0}^{\infty}\left(  2l+1\right)  P_{l}\left(  \cos\theta\right)
P_{l}\left(  \cos\theta^{\prime}\right)  =2\delta\left(  \cos\theta-\cos
\theta^{\prime}\right)
\end{equation}
and performing the integral, we have%
\begin{align}
\Sigma_{1}\left(  k;r,r^{\prime}\right)   &  =-\frac{1}{2}\int_{0}^{\pi}%
\frac{\cos qr}{qr}\mathrm{d}\cos\theta\int_{0}^{\pi}\mathrm{d}\cos
\theta^{\prime}\frac{\sin q^{\prime}r^{\prime}}{q^{\prime}r^{\prime}}%
\delta\left(  \cos\theta-\cos\theta^{\prime}\right) \nonumber\\
&  =-\frac{1}{2}\int_{0}^{\pi}\frac{\cos qr}{qr}\frac{\sin qr^{\prime}%
}{qr^{\prime}}\mathrm{d}\cos\theta\nonumber\\
&  =\frac{\operatorname*{Si}\left(  2kr-2kr^{\prime}\right)
-\operatorname*{Si}\left(  2kr+2kr^{\prime}\right)  }{4k^{2}rr^{\prime}},
\label{w1}%
\end{align}
where $\operatorname*{Si}\left(  z\right)  $ is the Sine integral function.

Similarly, we obtain%
\begin{equation}
\Sigma_{2}\left(  k;r,r^{\prime}\right)  =-\frac{\operatorname*{Si}\left(
2kr+2kr^{\prime}\right)  +\operatorname*{Si}\left(  2kr-2kr^{\prime}\right)
}{4k^{2}rr^{\prime}}. \label{w2}%
\end{equation}

Substituting Eqs. (\ref{w1}) and (\ref{w2}) into Eq. (\ref{kb2}) gives the
second-order global heat kernel:
\begin{align}
K^{\text{s}\left(  2\right)  }\left(  t\right)   &  =-\frac{t}{4\pi}\int%
_{0}^{\infty}V\left(  r\right)  r\mathrm{d}r\int_{0}^{r}V\left(  r^{\prime
}\right)  r^{\prime}\mathrm{d}r^{\prime}\int_{0}^{\infty}\mathrm{d}%
k^{2}\mathrm{e}^{-k^{2}t}\left[  \operatorname*{Si}\left(  2kr-2kr^{\prime
}\right)  -\operatorname*{Si}\left(  2kr+2kr^{\prime}\right)  \right]
\nonumber\\
&  +\frac{t}{4\pi}\int_{0}^{\infty}V\left(  r\right)  r\mathrm{d}r\int%
_{r}^{\infty}V\left(  r^{\prime}\right)  r^{\prime}\mathrm{d}r^{\prime}%
\int_{0}^{\infty}\mathrm{d}k^{2}\mathrm{e}^{-k^{2}t}\left[  \operatorname*{Si}%
\left(  2kr+2kr^{\prime}\right)  +\operatorname*{Si}\left(  2kr-2kr^{\prime
}\right)  \right]  .
\end{align}
The integral of $k^{2}$ is a Laplace transform. Performing the Laplace
transform gives the second-order Born approximation of heat kernels:%
\begin{align}
K^{\text{s}\left(  2\right)  }\left(  t\right)   &  =-\frac{1}{8}\int%
_{0}^{\infty}V\left(  r\right)  r\mathrm{d}r\left\{  \int_{0}^{r}%
\mathrm{d}r^{\prime}r^{\prime}V\left(  r^{\prime}\right)  \left[
\operatorname*{erf}\left(  \frac{r-r^{\prime}}{\sqrt{t}}\right)
-\operatorname*{erf}\left(  \frac{r+r^{\prime}}{\sqrt{t}}\right)  \right]
\right. \nonumber\\
&  -\left.  \int_{r}^{\infty}\mathrm{d}r^{\prime}V\left(  r^{\prime}\right)
r^{\prime}\left[  \operatorname*{erf}\left(  \frac{r^{\prime}-r}{\sqrt{t}%
}\right)  +\operatorname*{erf}\left(  \frac{r+r^{\prime}}{\sqrt{t}}\right)
\right]  \right\}  , \label{ktborn2}%
\end{align}
where $\operatorname*{erf}\left(  z\right)  $ is the Error function
\cite{olver2010nist}.

\subsubsection{One-loop effective action}

We next calculate the second-order Born approximation of the one-loop
effective action by directly using the relation between heat kernels and
one-loop effective actions, Eq. (\ref{wsandrehe}),\ though, in principle, we
can also obtain this result by substituting the second-order phase shift
(\ref{born2}) into Eq. (\ref{1loop}).

Substituting Eq. (\ref{ktborn2}) into Eq. (\ref{wsandrehe}) gives the
second-order Born approximation of the one-loop effective action:%
\begin{align}
W^{\left(  2\right)  }\left(  s\right)  =  &  \frac{1}{16}\tilde{\mu}^{2s}%
\int_{0}^{\infty}\mathrm{d}rrV\left(  r\right)  \left\{  \int_{0}%
^{r}\mathrm{d}r^{\prime}r^{\prime}V\left(  r^{\prime}\right)  \int_{0}%
^{\infty}\left[  \operatorname*{erf}\left(  \frac{r-r^{\prime}}{\sqrt{t}%
}\right)  -\operatorname*{erf}\left(  \frac{r+r^{\prime}}{\sqrt{t}}\right)
\right]  t^{s-1}\mathrm{d}t\right. \nonumber\\
&  \left.  -\int_{r}^{\infty}\mathrm{d}r^{\prime}r^{\prime}V\left(  r^{\prime
}\right)  \int_{0}^{\infty}\left[  \operatorname*{erf}\left(  \frac{r^{\prime
}-r}{\sqrt{t}}\right)  +\operatorname*{erf}\left(  \frac{r+r^{\prime}}%
{\sqrt{t}}\right)  \right]  t^{s-1}\mathrm{d}t\right\}  .
\end{align}
Performing the integral, we have%
\begin{align}
W^{\left(  2\right)  }\left(  s\right)  =  &  \frac{\tilde{\mu}^{2s}%
\Gamma\left(  \frac{1}{2}-s\right)  }{16\sqrt{\pi}s}\int_{0}^{\infty
}\mathrm{d}rrV\left(  r\right)  \left\{  \int_{0}^{r}\mathrm{d}r^{\prime
}r^{\prime}V\left(  r^{\prime}\right)  \left[  \left(  r-r^{\prime}\right)
^{2s}-\left(  r+r^{\prime}\right)  ^{2s}\right]  \right. \nonumber\\
&  \left.  -\int_{r}^{\infty}\mathrm{d}r^{\prime}r^{\prime}V\left(  r^{\prime
}\right)  \left[  \left(  r-r^{\prime}\right)  ^{2s}+\left(  r+r^{\prime
}\right)  ^{2s}\right]  \right\}  .
\end{align}

\subsubsection{Vacuum energy}

Similarly, using the relation between the heat kernel and the vacuum energy,
Eq. (\ref{eandrehe}), we can obtain the second-order Born approximation of the
vacuum energy. Substituting Eq. (\ref{ktborn2}) into Eq. (\ref{eandrehe})
gives%
\begin{align}
E_{0}^{\left(  2\right)  }\left(  \epsilon\right)   &  =-\frac{\mu^{2\epsilon
}\Gamma\left(  1-\epsilon\right)  }{16\sqrt{\pi}\Gamma\left(  \epsilon
+\frac{1}{2}\right)  }\int_{0}^{\infty}\mathrm{d}rrV\left(  r\right)  \left\{
\int_{0}^{r}\mathrm{d}r^{\prime}r^{\prime}V\left(  r^{\prime}\right)  \left[
\left(  r-r^{\prime}\right)  ^{2\epsilon-1}+\left(  r+r^{\prime}\right)
^{2\epsilon-1}\right]  \right. \nonumber\\
&  \left.  -\int_{r}^{\infty}\mathrm{d}r^{\prime}r^{\prime}V\left(  r^{\prime
}\right)  \left[  \left(  r-r^{\prime}\right)  ^{2\epsilon-1}+\left(
r+r^{\prime}\right)  ^{2\epsilon-1}\right]  \right\}  .
\end{align}

\section{Born approximation: $n$-dimensional case \label{BAnD}}

In this section, we consider the $n$-dimensional Born approximation. The
$n$-dimensional results can be used\ to perform the dimensional
renormalization. The dimensional renormalization can remove the divergence in
the Born approximation \cite{li2016scattering}.

\subsection{First-order Born approximation}

The first-order Born approximation of scattering phase shifts in $n$
dimensions is \cite{graham2009spectral}%
\begin{equation}
\delta_{l}^{\left(  1\right)  }\left(  k\right)  =-\frac{\pi}{2}\int%
_{0}^{\infty}J_{\frac{n}{2}+l-1}^{2}\left(  kr\right)  V\left(  r\right)
r\mathrm{d}r. \label{phaseshiftnd}%
\end{equation}

\subsubsection{Heat kernel}

The first-order approximation of the heat kernel can be obtained by
substituting Eq. (\ref{phaseshiftnd}) into Eq. (\ref{psdaorehe}):%
\begin{equation}
K^{\text{s}\left(  1\right)  }\left(  t\right)  =-\frac{t}{2}\int_{0}^{\infty
}V\left(  r\right)  r\mathrm{d}r\int_{0}^{\infty}\mathrm{d}k^{2}%
\mathrm{e}^{-k^{2}t}\sum_{l=0}^{\infty}D_{l}J_{\frac{n}{2}+l-1}^{2}\left(
kr\right)  . \label{klsb11}%
\end{equation}
The $n$-dimensional degeneracy for spherically symmetric potentials is
\cite{graham2009spectral}%
\begin{equation}
D_{l}=\frac{\left(  n+2l-2\right)  \Gamma\left(  n+l-2\right)  }{\Gamma\left(
n-1\right)  \Gamma\left(  l+1\right)  }.
\end{equation}
Taking $q=\frac{n}{2}-1$ in the sum rule \cite{graham2009spectral},%
\begin{equation}
\sum_{l=0}^{\infty}\frac{\left(  2q+2l\right)  \Gamma\left(  2q+l\right)
}{\Gamma\left(  l+1\right)  }J_{q+l}^{2}\left(  z\right)  =\frac{\Gamma\left(
2q+1\right)  }{\Gamma\left(  q+1\right)  ^{2}}\left(  \frac{z}{2}\right)
^{2q},
\end{equation}
gives%
\begin{equation}
\sum_{l=0}^{\infty}\frac{\Gamma\left(  n+l-2\right)  \left(  n+2l-2\right)
}{\Gamma\left(  n-1\right)  \Gamma\left(  l+1\right)  }J_{\frac{n}{2}+l-1}%
^{2}\left(  kr\right)  =\frac{1}{\Gamma\left(  \frac{n}{2}\right)  ^{2}%
}\left(  \frac{kr}{2}\right)  ^{n-2}. \label{sum2}%
\end{equation}
Substituting Eq. (\ref{sum2}) into Eq. (\ref{klsb11}) gives the first-order
Born approximation of the heat kernel:%
\begin{equation}
K^{\text{s}\left(  1\right)  }\left(  t\right)  =-\frac{t}{2}\frac{1}%
{\Gamma\left(  \frac{n}{2}\right)  ^{2}}\int_{0}^{\infty}V\left(  r\right)
r\mathrm{d}r\int_{0}^{\infty}\mathrm{d}k^{2}\mathrm{e}^{-k^{2}t}\left(
\frac{kr}{2}\right)  ^{n-2}.\nonumber
\end{equation}
Performing the Laplace transform, we arrive at the first-order Born
approximation of the $n$-dimensional heat kernel:%
\begin{equation}
K^{\text{s}\left(  1\right)  }\left(  t\right)  =-\frac{2^{1-n}}{\Gamma\left(
\frac{n}{2}\right)  t^{n/2-1}}\int_{0}^{\infty}V\left(  r\right)
r^{n-1}\mathrm{d}r.
\end{equation}

\subsubsection{One-loop effective action}

The first-order $n$-dimensional one-loop effective action can be obtained by
substituting the first-order $n$-dimensional phase shift (\ref{phaseshiftnd})
into Eq. (\ref{1loop}):%
\begin{equation}
W^{\left(  1\right)  }\left(  s\right)  =\frac{1}{4}\tilde{\mu}^{2s}%
\Gamma\left(  s+1\right)  \int_{0}^{\infty}V\left(  r\right)  r\mathrm{d}%
r\int_{0}^{\infty}\mathrm{d}k^{2}\left(  k^{2}\right)  ^{-s-1}\sum
_{l=0}^{\infty}D_{l}J_{\frac{n}{2}+l-1}^{2}\left(  kr\right)  .
\end{equation}
Using the sum rule (\ref{sum2}), we arrive at%
\begin{equation}
W^{\left(  1\right)  }\left(  s\right)  =\frac{1}{2^{n}\Gamma\left(  \frac
{n}{2}\right)  ^{2}}\tilde{\mu}^{2s}\Gamma\left(  s+1\right)  \int_{0}%
^{\infty}V\left(  r\right)  r^{n-1}\mathrm{d}r\int_{0}^{\infty}\mathrm{d}%
k^{2}\left(  k^{2}\right)  ^{-s-1}k^{n-2}.
\end{equation}
Here the integral of $k$ may diverge. According to Ref.
\cite{vassilevich2005heat}, we rewrite $\left(  k^{2}\right)  ^{-s-1}$ as
$\left(  k^{2}+m^{2}\right)  ^{-s-1}$:%
\begin{equation}
W^{\left(  1\right)  }\left(  s\right)  =\frac{1}{2^{n}\Gamma\left(  \frac
{n}{2}\right)  ^{2}}\tilde{\mu}^{2s}\Gamma\left(  s+1\right)  \int_{0}%
^{\infty}V\left(  r\right)  r^{n-1}\mathrm{d}r\int_{0}^{\infty}\mathrm{d}%
k^{2}\left(  k^{2}+m^{2}\right)  ^{-s-1}k^{n-2}.
\end{equation}
Working out the integral, we obtain the first-order Born approximation of the
one-loop effective action:%
\begin{equation}
W^{\left(  1\right)  }\left(  s\right)  =\frac{\tilde{\mu}^{2s}\Gamma\left(
s+1-\frac{n}{2}\right)  }{2^{n}\Gamma\left(  \frac{n}{2}\right)  }\left(
m^{2}\right)  ^{\frac{n}{2}-1-s}\int_{0}^{\infty}V\left(  r\right)
r^{n-1}\mathrm{d}r.
\end{equation}

\subsubsection{Vacuum energy}

The first-order $n$-dimensional vacuum energy can be obtained by substituting
the first-order $n$-dimensional phase shift (\ref{phaseshiftnd}) into Eq.
(\ref{vE}):%
\begin{equation}
E_{0}^{\left(  1\right)  }\left(  \epsilon\right)  =-\frac{1}{4}\tilde{\mu
}^{2\epsilon}\frac{\Gamma\left(  \frac{1}{2}+\epsilon\right)  }{\Gamma\left(
-\frac{1}{2}+\epsilon\right)  }\int_{0}^{\infty}V\left(  r\right)
r\mathrm{d}r\int_{0}^{\infty}\mathrm{d}k^{2}\left(  k^{2}\right)
^{-1/2-\epsilon}\sum_{l=0}^{\infty}D_{l}J_{\frac{n}{2}+l-1}^{2}\left(
kr\right)  .
\end{equation}
Using the sum rule (\ref{sum2}), we arrive at%
\begin{equation}
E_{0}^{\left(  1\right)  }\left(  \epsilon\right)  =-\frac{1}{2^{n}%
\Gamma\left(  \frac{n}{2}\right)  ^{2}}\tilde{\mu}^{2\epsilon}\frac
{\Gamma\left(  \frac{1}{2}+\epsilon\right)  }{\Gamma\left(  -\frac{1}%
{2}+\epsilon\right)  }\int_{0}^{\infty}V\left(  r\right)  r^{n-1}%
\mathrm{d}r\int_{0}^{\infty}\mathrm{d}k^{2}\left(  k^{2}\right)
^{-1/2-\epsilon}k^{n-2}.
\end{equation}
The integral of $k$ may diverge. According to Ref. \cite{vassilevich2005heat},
we rewrite $\left(  k^{2}\right)  ^{-1/2-\epsilon}$ as $\left(  k^{2}%
+m^{2}\right)  ^{-1/2-\epsilon}$:%
\begin{equation}
E_{0}^{\left(  1\right)  }\left(  \epsilon\right)  =-\frac{1}{2^{n}%
\Gamma\left(  \frac{n}{2}\right)  ^{2}}\tilde{\mu}^{2\epsilon}\frac
{\Gamma\left(  \frac{1}{2}+\epsilon\right)  }{\Gamma\left(  -\frac{1}%
{2}+\epsilon\right)  }\int_{0}^{\infty}V\left(  r\right)  r^{n-1}%
\mathrm{d}r\int_{0}^{\infty}\mathrm{d}k^{2}\left(  k^{2}+m^{2}\right)
^{-1/2-\epsilon}k^{n-2}.
\end{equation}
Working out the integral, we obtain the first-order Born approximation of the
$n$-dimensional vacuum energy:%
\begin{equation}
E_{0}^{\left(  1\right)  }\left(  \epsilon\right)  =-\frac{1}{2^{n}%
\Gamma\left(  \frac{n}{2}\right)  }\tilde{\mu}^{2\epsilon}\frac{\Gamma\left(
\epsilon+\frac{1}{2}-\frac{n}{2}\right)  }{\Gamma\left(  \epsilon-\frac{1}%
{2}\right)  }\left(  m^{2}\right)  ^{\frac{n}{2}-\frac{1}{2}-\epsilon}\int%
_{0}^{\infty}V\left(  r\right)  r^{n-1}\mathrm{d}r.
\end{equation}

In performing the integral in the Born approximation, one may encounter
divergences that do not come from the usual divergence in quantum field
theory. Such divergences can be removed by the procedure given in Ref.
\cite{li2016scattering}.

\subsection{Second-order Born approximation}

The $n$-dimensional second-order Born approximation of the scattering phase
shift for a spherically symmetric potential $V\left(  r\right)  $ is
\cite{joachain1975quantum}%
\begin{align}
\delta_{l}^{\left(  2\right)  } =  &  -\frac{\pi^{2}}{4}\int_{0}^{\infty
}J_{\frac{n}{2}+l-1}\left(  kr\right)  Y_{\frac{n}{2}+l-1}\left(  kr\right)
V\left(  r\right)  r\mathrm{d}r\int_{0}^{r}J_{\frac{n}{2}+l-1}^{2}\left(
kr^{\prime}\right)  V\left(  r^{\prime}\right)  r^{\prime}\mathrm{d}r^{\prime
}\nonumber\\
&  -\frac{\pi^{2}}{4}\int_{0}^{\infty}J_{\frac{n}{2}+l-1}^{2}\left(
kr\right)  V\left(  r\right)  r\mathrm{d}r\int_{r}^{\infty}J_{\frac{n}{2}%
+l-1}\left(  kr^{\prime}\right)  Y_{\frac{n}{2}+l-1}\left(  kr^{\prime
}\right)  V\left(  r^{\prime}\right)  r^{\prime}\mathrm{d}r^{\prime},
\label{ndborn2}%
\end{align}
where $Y_{\nu}\left(  z\right)  $ is the Bessel function of the second kind.

\subsubsection{Heat kernel}

The second-order approximation of the $n$-dimensional heat kernel can be
obtained by substituting the second-order phase shift\ (\ref{ndborn2}) into
Eq. (\ref{psdaorehe}):%
\begin{align}
K^{\text{s}\left(  2\right)  }\left(  t\right)  =  &  -\frac{\pi t}{4}\int%
_{0}^{\infty}\mathrm{e}^{-k^{2}t}\mathrm{d}k^{2}\int_{0}^{\infty}V\left(
r\right)  r\mathrm{d}r\int_{0}^{r}\mathrm{d}r^{\prime}V\left(  r^{\prime
}\right)  r^{\prime}\Sigma_{1}\left(  k;r,r^{\prime}\right) \nonumber\\
&  -\frac{\pi t}{4}\int_{0}^{\infty}\mathrm{e}^{-k^{2}t}\mathrm{d}k^{2}%
\int_{0}^{\infty}V\left(  r\right)  r\mathrm{d}r\int_{r}^{\infty}%
\mathrm{d}r^{\prime}V\left(  r^{\prime}\right)  r^{\prime}\Sigma_{2}\left(
k;r,r^{\prime}\right)  , \label{kb2nd}%
\end{align}
where%
\begin{align}
\Sigma_{1}\left(  k;r,r^{\prime}\right)   &  =\sum_{l=0}^{\infty}\frac{\left(
n+2l-2\right)  \Gamma\left(  n+l-2\right)  }{\Gamma\left(  n-1\right)
\Gamma\left(  l+1\right)  }J_{\frac{n}{2}+l-1}\left(  kr\right)  Y_{\frac
{n}{2}+l-1}\left(  kr\right)  J_{\frac{n}{2}+l-1}^{2}\left(  kr^{\prime
}\right)  ,\label{ndw1}\\
\Sigma_{2}\left(  k;r,r^{\prime}\right)   &  =\sum_{l=0}^{\infty}\frac{\left(
n+2l-2\right)  \Gamma\left(  n+l-2\right)  }{\Gamma\left(  n-1\right)
\Gamma\left(  l+1\right)  }J_{\frac{n}{2}+l-1}^{2}\left(  kr\right)
J_{\frac{n}{2}+l-1}\left(  kr^{\prime}\right)  Y_{\frac{n}{2}+l-1}\left(
kr^{\prime}\right)  . \label{ndw2}%
\end{align}

To perform these sums, we give an integral representation of $J_{l+\mu}%
^{2}\left(  kr\right)  $ in Appendix \ref{Appendix3},%
\begin{equation}
J_{l+\mu}^{2}\left(  kr\right)  =\frac{2^{\mu-1}\Gamma\left(  l+1\right)
\Gamma\left(  \mu\right)  }{\pi\Gamma\left(  2\mu+l\right)  }\left(
kr\right)  ^{2\mu}\int_{0}^{\pi}\frac{J_{\mu}\left(  qr\right)  }{q^{\mu
}r^{\mu}}C_{l}^{\mu}\left(  \cos\theta\right)  \sin^{2\mu}\theta
\mathrm{d}\theta,
\end{equation}
and an integral representation of $J_{l+\mu}\left(  kr\right)  Y_{l+\mu
}\left(  kr\right)  $ in Appendix \ref{Appendix4},%
\begin{equation}
J_{l+\mu}\left(  kr\right)  Y_{l+\mu}\left(  kr\right)  =\frac{2^{\mu-1}%
\Gamma\left(  l+1\right)  \Gamma\left(  \mu\right)  }{\pi\Gamma\left(
2\mu+l\right)  }\left(  kr\right)  ^{2\mu}\int_{0}^{\pi}\frac{Y_{\mu}\left(
qr\right)  }{q^{\mu}r^{\mu}}C_{l}^{\mu}\left(  \cos\theta\right)  \sin^{2\mu
}\theta\mathrm{d}\theta,
\end{equation}
where $C_{l}^{\mu}\left(  \cos\theta\right)  $ is the Gegenbauer polynomial.
Substituting the above two integral representations into Eq. (\ref{ndw1})
gives%
\begin{align}
\Sigma_{1}\left(  k;r,r^{\prime}\right)   &  =\sum_{l=0}^{\infty}\frac{\left(
2\mu+2l\right)  \Gamma\left(  2\mu+l\right)  }{\Gamma\left(  2\mu+1\right)
\Gamma\left(  l+1\right)  }J_{l+\mu}\left(  kr\right)  Y_{l+\mu}\left(
kr\right)  J_{l+\mu}^{2}\left(  kr^{\prime}\right) \nonumber\\
&  =\sum_{l=0}^{\infty}\frac{\left(  2\mu+2l\right)  \Gamma\left(
2\mu+l\right)  }{\Gamma\left(  2\mu+1\right)  \Gamma\left(  l+1\right)
}\left[  \frac{2^{\mu-1}\Gamma\left(  l+1\right)  \Gamma\left(  \mu\right)
}{\pi\Gamma\left(  2\mu+l\right)  }\left(  kr\right)  ^{2\mu}\int_{0}^{\pi
}\frac{Y_{\mu}\left(  qr\right)  }{\left(  qr\right)  ^{\mu}}C_{l}^{\mu
}\left(  \cos\theta\right)  \sin^{2\mu}\theta\mathrm{d}\theta\right.
\nonumber\\
&  \left.  \times\frac{2^{\mu-1}\Gamma\left(  l+1\right)  \Gamma\left(
\mu\right)  }{\pi\Gamma\left(  2\mu+l\right)  }\left(  kr^{\prime}\right)
^{2\mu}\int_{0}^{\pi}\frac{J_{\mu}\left(  q^{\prime}r^{\prime}\right)
}{\left(  q^{\prime}r^{\prime}\right)  ^{\mu}}C_{l}^{\mu}\left(  \cos
\theta^{\prime}\right)  \sin^{2\mu}\theta^{\prime}\mathrm{d}\theta^{\prime
}\right] \nonumber\\
&  =\frac{2^{2\left(  \mu-1\right)  }\Gamma^{2}\left(  \mu\right)  }{\pi
^{2}\Gamma\left(  2\mu+1\right)  }\left(  kr\right)  ^{2\mu}\left(
kr^{\prime}\right)  ^{2\mu}\int_{0}^{\pi}\mathrm{d}\theta\frac{Y_{\mu}\left(
qr\right)  }{\left(  qr\right)  ^{\mu}}\sin^{2\mu}\theta\left(  kr^{\prime
}\right)  ^{2\mu}\int_{0}^{\pi}\mathrm{d}\theta^{\prime}\frac{J_{\mu}\left(
q^{\prime}r^{\prime}\right)  }{\left(  q^{\prime}r^{\prime}\right)  ^{\mu}%
}\sin^{2\mu}\theta^{\prime}\nonumber\\
&  \times\sum_{l=0}^{\infty}\frac{\left(  2\mu+2l\right)  \Gamma\left(
l+1\right)  }{\Gamma\left(  2\mu+l\right)  }C_{l}^{\mu}\left(  \cos
\theta\right)  C_{l}^{\mu}\left(  \cos\theta^{\prime}\right)  ,
\end{align}
where $\mu=\frac{n}{2}-1$, $q=2k\sin\frac{\theta}{2}$, and $q^{\prime}%
=2k\sin\frac{\theta^{\prime}}{2}$. Using the relation \cite{olver2010nist}%
\begin{equation}
\sum_{l=0}^{\infty}\frac{\Gamma\left(  l+1\right)  \left(  2\mu+2l\right)
}{\Gamma\left(  2\mu+l\right)  }C_{l}^{\mu}\left(  \cos\theta\right)
C_{l}^{\mu}\left(  \cos\theta^{\prime}\right)  =\frac{2^{2-2\mu}\pi}%
{\Gamma^{2}\left(  \mu\right)  }\left(  \sin\theta\right)  ^{\frac{1-2\mu}{2}%
}\left(  \sin\theta^{\prime}\right)  ^{\frac{1-2\mu}{2}}\delta\left(
\cos\theta-\cos\theta^{\prime}\right)  ,
\end{equation}
we arrive at%
\begin{align}
\Sigma_{1}\left(  k;r,r^{\prime}\right)   &  =\frac{\left(  kr\right)  ^{2\mu
}\left(  kr^{\prime}\right)  ^{2\mu}}{\pi\Gamma\left(  2\mu+1\right)  }%
\int_{0}^{\pi}\frac{Y_{\mu}\left(  qr\right)  }{\left(  qr\right)  ^{\mu}}%
\sin^{\frac{2\mu-1}{2}}\theta\mathrm{d}\cos\theta\int_{0}^{\pi}\frac{J_{\mu
}\left(  qr^{\prime}\right)  }{\left(  q^{\prime}r^{\prime}\right)  ^{\mu}%
}\sin^{\frac{2\mu-1}{2}}\theta^{\prime}\delta\left(  \cos\theta-\cos
\theta^{\prime}\right)  \mathrm{d}\cos\theta^{\prime}\nonumber\\
&  =\frac{\left(  kr\right)  ^{2\mu}\left(  kr^{\prime}\right)  ^{2\mu}}%
{\pi\Gamma\left(  2\mu+1\right)  }\int_{0}^{\pi}\frac{Y_{\mu}\left(
qr\right)  }{\left(  qr\right)  ^{\mu}}\frac{J_{\mu}\left(  qr^{\prime
}\right)  }{\left(  qr^{\prime}\right)  ^{\mu}}\sin^{2\mu-1}\theta
\mathrm{d}\cos\theta. \label{w1nbw1}%
\end{align}

Similarly, we have%
\begin{equation}
\Sigma_{2}\left(  k;r,r^{\prime}\right)  =\frac{\left(  kr\right)  ^{2\mu
}\left(  kr^{\prime}\right)  ^{2\mu}}{\pi\Gamma\left(  2\mu+1\right)  }%
\int_{0}^{\pi}\frac{Y_{\mu}\left(  qr^{\prime}\right)  }{\left(  qr^{\prime
}\right)  ^{\mu}}\frac{J_{\mu}\left(  qr\right)  }{\left(  qr\right)  ^{\mu}%
}\sin^{2\mu-1}\theta\mathrm{d}\cos\theta. \label{w2ndw2}%
\end{equation}
Substituting Eqs. (\ref{w1nbw1}) and (\ref{w2ndw2}) into Eq. (\ref{kb2nd})
gives the $n$-dimensional second-order global heat kernel:%
\begin{align}
K^{\text{s}\left(  2\right)  }\left(  t\right)   &  =-\frac{t}{4\Gamma\left(
n-1\right)  }\int_{0}^{\infty}\mathrm{e}^{-k^{2}t}\mathrm{d}k^{2}\int%
_{0}^{\infty}\mathrm{d}rrV\left(  r\right)  \left(  kr\right)  ^{n-2}\left\{
\int_{0}^{r}\mathrm{d}r^{\prime}r^{\prime}V\left(  r^{\prime}\right)  \left[
\left(  kr^{\prime}\right)  ^{n-2}\int_{0}^{\pi}\frac{Y_{\frac{n}{2}-1}\left(
qr\right)  }{\left(  qr\right)  ^{\frac{n}{2}-1}}\frac{J_{\frac{n}{2}%
-1}\left(  qr^{\prime}\right)  }{\left(  qr^{\prime}\right)  ^{\frac{n}{2}-1}%
}\sin^{n-3}\theta\mathrm{d}\cos\theta\right]  \right. \nonumber\\
&  \left.  +\int_{r}^{\infty}\mathrm{d}r^{\prime}r^{\prime}V\left(  r^{\prime
}\right)  \left[  \left(  kr^{\prime}\right)  ^{n-2}\int_{0}^{\pi}%
\frac{Y_{\frac{n}{2}-1}\left(  qr^{\prime}\right)  }{\left(  qr^{\prime
}\right)  ^{\frac{n}{2}-1}}\frac{J_{\frac{n}{2}-1}\left(  qr\right)  }{\left(
qr\right)  ^{\frac{n}{2}-1}}\sin^{2\mu-1}\theta\mathrm{d}\cos\theta\right]
\right\}  . \label{ndkb2}%
\end{align}

The integral encountered in Eq. (\ref{ndkb2}) is difficult. The
odd-dimensional case and even-dimensional case are very different
\cite{milton2003calculating}, for the odd-dimensional Bessel polynomial is a
polynomial but the even-dimensional case is not \cite{li2016scattering}. In
the following we only consider the odd-dimensional case.

For odd-dimensional cases, the integral representation given in Appendix
\ref{Appendix4} with $\mu=\frac{1}{2}$ and $l=\frac{n}{2}-\frac{3}{2}$
($n=3,5,7,\cdots$), becomes
\begin{align}
Y_{\frac{n}{2}-1}\left(  qr\right)  J_{\frac{n}{2}-1}\left(  qr^{\prime
}\right)   &  =\frac{q\sqrt{rr^{\prime}}}{\sqrt{2\pi}}\int_{0}^{\pi}%
\frac{Y_{1/2}\left(  q\sqrt{r^{2}+r^{\prime2}-2rr^{\prime}\cos\phi}\right)
}{\left(  q\sqrt{r^{2}+r^{\prime2}-2rr^{\prime}\cos\phi}\right)  ^{1/2}%
}P_{\frac{n}{2}-\frac{3}{2}}\left(  \cos\phi\right)  \mathrm{d}\cos
\phi\nonumber\\
&  =-\frac{q\sqrt{rr^{\prime}}}{\pi}\int_{0}^{\pi}\frac{\cos\left(
q\sqrt{r^{2}+r^{\prime2}-2rr^{\prime}\cos\phi}\right)  }{q\sqrt{r^{2}%
+r^{\prime2}-2rr^{\prime}\cos\phi}}P_{\frac{n}{2}-\frac{3}{2}}\left(  \cos
\phi\right)  \mathrm{d}\cos\phi.
\end{align}
Then the integral over $\theta$ in Eq. (\ref{ndkb2}) reads
\begin{align}
&  \int_{0}^{\pi}\frac{Y_{\frac{n}{2}-1}\left(  2kr\sin\frac{\theta}%
{2}\right)  }{\left(  2kr\sin\frac{\theta}{2}\right)  ^{\frac{n}{2}-1}}%
\frac{J_{\frac{n}{2}-1}\left(  2kr^{\prime}\sin\frac{\theta}{2}\right)
}{\left(  2kr^{\prime}\sin\frac{\theta}{2}\right)  ^{\frac{n}{2}-1}}\sin
^{n-3}\theta\mathrm{d}\cos\theta\nonumber\\
&  =\int_{0}^{\pi}\frac{1}{\left(  2kr\sin\frac{\theta}{2}\right)  ^{\frac
{n}{2}-1}\left(  2kr^{\prime}\sin\frac{\theta}{2}\right)  ^{\frac{n}{2}-1}%
}\left[  -\frac{q\sqrt{rr^{\prime}}}{\pi}\right.  \left.  \int_{0}^{\pi}%
\frac{\cos\left(  q\sqrt{r^{2}+r^{\prime2}-2rr^{\prime}\cos\phi}\right)
}{q\sqrt{r^{2}+r^{\prime2}-2rr^{\prime}\cos\phi}}P_{\frac{n}{2}-\frac{3}{2}%
}\left(  \cos\phi\right)  \sin\phi\mathrm{d}\phi\right]  \sin^{n-3}%
\theta\mathrm{d}\cos\theta
\end{align}
with $q=2k\sin\frac{\theta}{2}$. We first perform the integral over $\theta$:%
\begin{equation}
\int_{0}^{\pi}\frac{\cos\left(  2k\sin\frac{\theta}{2}\sqrt{r^{2}+r^{\prime
2}-2rr^{\prime}\cos\phi}\right)  }{\left(  \sin\frac{\theta}{2}\right)
^{n-2}}\sin^{n-3}\theta\mathrm{d}\cos\theta=-2^{n-2}\sqrt{\pi}\Gamma\left(
\frac{n}{2}-\frac{1}{2}\right)  \frac{J_{\frac{n}{2}-1}\left(  2k\sqrt
{r^{2}+r^{\prime2}-2rr^{\prime}\cos\phi}\right)  }{\left(  k\sqrt
{r^{2}+r^{\prime2}-2rr^{\prime}\cos\phi}\right)  ^{\frac{n}{2}-1}}.
\end{equation}
Then we have%
\begin{equation}
\int_{0}^{\pi}\frac{Y_{\frac{n}{2}-1}\left(  qr\right)  }{\left(  qr\right)
^{\frac{n}{2}-1}}\frac{J_{\frac{n}{2}-1}\left(  qr^{\prime}\right)  }{\left(
qr^{\prime}\right)  ^{\frac{n}{2}-1}}\sin^{n-3}\theta\mathrm{d}\cos
\theta=\frac{2^{\frac{n}{2}}\Gamma\left(  \frac{n}{2}-\frac{1}{2}\right)
}{k^{n-3}\sqrt{\pi}\left(  rr^{\prime}\right)  ^{\frac{n}{2}-\frac{3}{2}}}%
\int_{0}^{\pi}\frac{J_{\frac{n}{2}-1}\left(  2k\sqrt{r^{2}+r^{\prime
2}-2rr^{\prime}\cos\phi}\right)  }{\left(  2k\sqrt{r^{2}+r^{\prime
2}-2rr^{\prime}\cos\phi}\right)  ^{\frac{n}{2}}}P_{\frac{n}{2}-\frac{3}{2}%
}\left(  \cos\phi\right)  \mathrm{d}\cos\theta. \label{int1}%
\end{equation}

Eq. (\ref{ndkb2}), by Eq. (\ref{int1}), becomes%
\begin{align}
K^{\text{s}\left(  2\right)  }\left(  t\right)   &  =-\frac{t}{4\sqrt{\pi}%
}\frac{\Gamma\left(  \frac{n}{2}-\frac{1}{2}\right)  }{\Gamma\left(
n-1\right)  }\int_{0}^{\infty}\mathrm{d}rr^{\frac{n+1}{2}}V\left(  r\right)
\nonumber\\
&  \times\left\{  \int_{0}^{r}\mathrm{d}r^{\prime}\left(  r^{\prime}\right)
^{\frac{n+1}{2}}V\left(  r^{\prime}\right)  \int_{0}^{\pi}\left[  \int%
_{0}^{\infty}\mathrm{e}^{-k^{2}t}k^{n-1}\frac{J_{\frac{n}{2}-1}\left(
2k\sqrt{r^{2}+r^{\prime2}-2rr^{\prime}\cos\phi}\right)  }{\left(  k\sqrt
{r^{2}+r^{\prime2}-2rr^{\prime}\cos\phi}\right)  ^{\frac{n}{2}}}%
\mathrm{d}k^{2}\right]  \right.  P_{\frac{n}{2}-\frac{3}{2}}\left(  \cos
\phi\right)  \mathrm{d}\cos\theta\nonumber\\
&  +\int_{r}^{\infty}\mathrm{d}r^{\prime}\left(  r^{\prime}\right)
^{\frac{n+1}{2}}V\left(  r^{\prime}\right)  \int_{0}^{\pi}\left[  \int%
_{0}^{\infty}\mathrm{e}^{-k^{2}t}k^{n-1}\frac{J_{\frac{n}{2}-1}\left(
2k\sqrt{r^{2}+r^{\prime2}-2rr^{\prime}\cos\phi}\right)  }{\left(  k\sqrt
{r^{2}+r^{\prime2}-2rr^{\prime}\cos\phi}\right)  ^{\frac{n}{2}}}%
\mathrm{d}k^{2}\right]  \left.  P_{\frac{n}{2}-\frac{3}{2}}\left(  \cos
\phi\right)  \mathrm{d}\cos\theta\right\}  .
\end{align}
Performing the integral over $k$, which is a Laplace transform, gives%
\begin{align}
K^{\text{s}\left(  2\right)  }\left(  t\right)   &  =-\frac{1}{2\sqrt{2}\pi
t^{\frac{n}{2}-\frac{1}{2}}}\frac{\Gamma\left(  \frac{n}{2}-\frac{1}%
{2}\right)  }{\Gamma\left(  n-1\right)  }\nonumber\\
&  \times\int_{0}^{\infty}r^{\frac{n+1}{2}}V\left(  r\right)  \mathrm{d}%
r\left\{  \int_{0}^{r}\mathrm{d}r^{\prime}\left(  r^{\prime}\right)
^{\frac{n+1}{2}}V\left(  r^{\prime}\right)  \int_{0}^{\pi}K_{1/2}\left(
\frac{r^{2}+r^{\prime2}-2rr^{\prime}\cos\phi}{t}\right)  P_{\frac{n}{2}%
-\frac{3}{2}}\left(  \cos\phi\right)  \mathrm{d}\cos\phi\right. \\
&  \left.  +\int_{r}^{\infty}\mathrm{d}r^{\prime}\left(  r^{\prime}\right)
^{\frac{n+1}{2}}V\left(  r^{\prime}\right)  \int_{0}^{\pi}K_{1/2}\left(
\frac{r^{2}+r^{\prime2}-2rr^{\prime}\cos\phi}{t}\right)  P_{\frac{n}{2}%
-\frac{3}{2}}\left(  \cos\phi\right)  \mathrm{d}\cos\phi\right\}  .
\end{align}
By $\sqrt{r^{2}+r^{\prime2}-2rr^{\prime}\cos\phi}=\left\vert \mathbf{r}%
-\mathbf{r}^{\prime}\right\vert $, we rewrite%
\begin{align}
K^{\text{s}\left(  2\right)  }\left(  t\right)   &  =-\frac{1}{2\sqrt{2}\pi
t^{\frac{n}{2}-\frac{1}{2}}}\frac{\Gamma\left(  \frac{n}{2}-\frac{1}%
{2}\right)  }{\Gamma\left(  n-1\right)  }\int_{0}^{\infty}r^{\frac{n+1}{2}%
}V\left(  r\right)  \mathrm{d}r\left\{  \int_{0}^{r}\mathrm{d}r^{\prime
}\left(  r^{\prime}\right)  ^{\frac{n+1}{2}}V\left(  r^{\prime}\right)
\int_{0}^{\pi}K_{1/2}\left(  \frac{\left(  \mathbf{r}-\mathbf{r}^{\prime
}\right)  ^{2}}{t}\right)  P_{\frac{n}{2}-\frac{3}{2}}\left(  \cos\phi\right)
\mathrm{d}\cos\phi\right. \nonumber\\
&  \left.  +\int_{r}^{\infty}\mathrm{d}r^{\prime}\left(  r^{\prime}\right)
^{\frac{n+1}{2}}V\left(  r^{\prime}\right)  \int_{0}^{\pi}K_{1/2}\left(
\frac{\left(  \mathbf{r}-\mathbf{r}^{\prime}\right)  ^{2}}{t}\right)
P_{\frac{n}{2}-\frac{3}{2}}\left(  \cos\phi\right)  \mathrm{d}\cos
\phi\right\}  , \label{nk2}%
\end{align}
or, equivalently,%
\begin{align}
K^{\text{s}\left(  2\right)  }\left(  t\right)   &  =-\frac{1}{4\sqrt{\pi
}t^{\frac{n}{2}-1}}\frac{\Gamma\left(  \frac{n}{2}-\frac{1}{2}\right)
}{\Gamma\left(  n-1\right)  }\int_{0}^{\infty}r^{\frac{n+1}{2}}V\left(
r\right)  \mathrm{d}r\left\{  \int_{0}^{r}\mathrm{d}r^{\prime}\left(
r^{\prime}\right)  ^{\frac{n+1}{2}}V\left(  r^{\prime}\right)  \int_{0}^{\pi
}\frac{\exp\left(  -\frac{\left(  \mathbf{r}-\mathbf{r}^{\prime}\right)  ^{2}%
}{t}\right)  }{\left\vert \mathbf{r}-\mathbf{r}^{\prime}\right\vert }%
P_{\frac{n}{2}-\frac{3}{2}}\left(  \cos\phi\right)  \mathrm{d}\cos\phi\right.
\nonumber\\
&  \left.  +\int_{r}^{\infty}\mathrm{d}r^{\prime}\left(  r^{\prime}\right)
^{\frac{n+1}{2}}V\left(  r^{\prime}\right)  \int_{0}^{\pi}\frac{\exp\left(
-\frac{\left(  \mathbf{r}-\mathbf{r}^{\prime}\right)  ^{2}}{t}\right)
}{\left\vert \mathbf{r}-\mathbf{r}^{\prime}\right\vert }P_{\frac{n}{2}%
-\frac{3}{2}}\left(  \cos\phi\right)  \mathrm{d}\cos\phi\right\}  ,
\label{ndkb2222}%
\end{align}
where $K_{\nu}\left(  z\right)  $ is the modified Bessel function of the
second kind \cite{olver2010nist}.

In the heat-kernel theory, one often concentrates on the small $t$ case, e.g.,
the Seeley-DeWitt expansion \cite{vassilevich2003heat}. For small $t$, we have
the following expansion:%
\begin{equation}
K_{1/2}\left(  \frac{\left(  \mathbf{r}-\mathbf{r}^{\prime}\right)  ^{2}}%
{t}\right)  \sim\sqrt{\frac{\pi}{2}}\frac{1}{\sqrt{r^{2}+r^{\prime
2}-2rr^{\prime}\cos\phi}}+\cdots.
\end{equation}
Substituting into Eq. (\ref{nk2}) gives%
\begin{align}
K^{\text{s}\left(  2\right)  }\left(  t\right)   &  \sim-\frac{1}{4\sqrt{\pi
}t^{\frac{n}{2}-1}}\frac{\Gamma\left(  \frac{n}{2}-\frac{1}{2}\right)
}{\Gamma\left(  n-1\right)  }\int_{0}^{\infty}r^{\frac{n+1}{2}}V\left(
r\right)  \mathrm{d}r\left\{  \int_{0}^{r}\mathrm{d}r^{\prime}\left(
r^{\prime}\right)  ^{\frac{n+1}{2}}V\left(  r^{\prime}\right)  \int_{0}^{\pi
}\frac{1}{\sqrt{r^{2}+r^{\prime2}-2rr^{\prime}\cos\phi}}P_{\frac{n}{2}%
-\frac{3}{2}}\left(  \cos\phi\right)  \mathrm{d}\cos\phi\right. \nonumber\\
&  \left.  +\int_{r}^{\infty}\mathrm{d}r^{\prime}\left(  r^{\prime}\right)
^{\frac{n+1}{2}}V\left(  r^{\prime}\right)  \int_{0}^{\pi}\frac{1}{\sqrt
{r^{2}+r^{\prime2}-2rr^{\prime}\cos\phi}}P_{\frac{n}{2}-\frac{3}{2}}\left(
\cos\phi\right)  \mathrm{d}\cos\phi\right\}  . \label{Ks2t}%
\end{align}
To perform the integral in Eq. (\ref{Ks2t}), we use \cite{olver2010nist}%
\begin{equation}
\left\{
\begin{array}
[c]{c}%
\displaystyle\frac{1}{\sqrt{r^{2}+r^{\prime2}-2rr^{\prime}\cos\phi}}=\frac
{1}{r}\sum_{l=0}^{\infty}\left(  \frac{r^{\prime}}{r}\right)  ^{l}P_{l}\left(
\cos\phi\right)  ,\text{ \ }r>r^{\prime},\\
\displaystyle\frac{1}{\sqrt{r^{2}+r^{\prime2}-2rr^{\prime}\cos\phi}}=\frac
{1}{r^{\prime}}\sum_{l=0}^{\infty}\left(  \frac{r}{r^{\prime}}\right)
^{l}P_{l}\left(  \cos\phi\right)  ,\text{ \ }r<r^{\prime}.
\end{array}
\right.
\end{equation}
In Eq. (\ref{Ks2t}), the first term\ corresponds to $r>r^{\prime}$ and the
second term corresponds to $r<r^{\prime}$. Then
\begin{align}
\int_{0}^{\pi}\frac{1}{\sqrt{r^{2}+r^{\prime2}-2rr^{\prime}\cos\phi}}%
P_{\frac{n}{2}-\frac{3}{2}}\left(  \cos\phi\right)  \mathrm{d}\cos\phi &
=\frac{1}{r}\sum_{l=0}^{\infty}\left(  \frac{r^{\prime}}{r}\right)  ^{l}%
\int_{0}^{\pi}P_{l}\left(  \cos\phi\right)  P_{\frac{n}{2}-\frac{3}{2}}\left(
\cos\phi\right)  \mathrm{d}\cos\phi\nonumber\\
&  =\frac{1}{r}\sum_{l=0}^{\infty}\left(  \frac{r^{\prime}}{r}\right)
^{l}\frac{2}{2l+1}\delta_{l,\frac{n}{2}-\frac{3}{2}}\nonumber\\
&  =\frac{2}{n-2}\frac{1}{r}\left(  \frac{r^{\prime}}{r}\right)  ^{\frac{n}%
{2}-\frac{3}{2}},\text{ \ }\left(  r>r^{\prime}\right)  .
\end{align}
Similarly,%
\begin{equation}
\int_{0}^{\pi}\frac{1}{\sqrt{r^{2}+r^{\prime2}-2rr^{\prime}\cos\phi}}%
P_{\frac{n}{2}-\frac{3}{2}}\left(  \cos\phi\right)  \mathrm{d}\cos\phi
=\frac{2}{n-2}\frac{1}{r^{\prime}}\left(  \frac{r}{r^{\prime}}\right)
^{\frac{n}{2}-\frac{3}{2}},\text{ \ }\left(  r<r^{\prime}\right)  .
\end{equation}
The second-order heat kernel in odd dimensions, by Eq. (\ref{Ks2t}), reads
\begin{equation}
K^{\text{s}\left(  2\right)  }\left(  t\right)  \sim-\frac{1}{2\left(
n-2\right)  \sqrt{\pi}t^{\frac{n}{2}-1}}\frac{\Gamma\left(  \frac{n}{2}%
-\frac{1}{2}\right)  }{\Gamma\left(  n-1\right)  }\left[  \int_{0}^{\infty
}\mathrm{d}rrV\left(  r\right)  \int_{0}^{r}\mathrm{d}r^{\prime}\left(
r^{\prime}\right)  ^{n-1}V\left(  r^{\prime}\right)  +\int_{0}^{\infty
}\mathrm{d}rr^{n-1}V\left(  r\right)  \int_{r}^{\infty}\mathrm{d}r^{\prime
}r^{\prime}V\left(  r^{\prime}\right)  \right]  .
\end{equation}

\subsubsection{One-loop effective action}

Next by using the relation between the global heat kernel and the one-loop
effective action, Eq. (\ref{wsandrehe}), and substituting Eq. (\ref{ndkb2}%
)\ into Eq. (\ref{wsandrehe}), we obtain the second-order $n$-dimensional
one-loop effective action:%
\begin{align}
W^{\left(  2\right)  }\left(  s\right)   &  =\frac{\pi}{8}\tilde{\mu}^{2s}%
\int_{0}^{\infty}V\left(  r\right)  r\mathrm{d}r\nonumber\\
&  \times\left\{  \int_{0}^{r}V\left(  r^{\prime}\right)  r^{\prime}%
\mathrm{d}r^{\prime}\int_{0}^{\infty}\mathrm{d}k^{2}\left[  \frac{\left(
kr\right)  ^{n-2}\left(  kr^{\prime}\right)  ^{n-2}}{\pi\Gamma\left(
2\mu+1\right)  }\int_{0}^{\pi}\frac{Y_{\frac{n}{2}-1}\left(  qr\right)
}{\left(  qr\right)  ^{\frac{n}{2}-1}}\frac{J_{\frac{n}{2}-1}\left(
qr^{\prime}\right)  }{\left(  qr^{\prime}\right)  ^{\frac{n}{2}-1}}\sin
^{n-3}\theta d\cos\theta\right]  \right.  \int_{0}^{\infty}\mathrm{e}%
^{-k^{2}t}t^{s}\mathrm{d}t\nonumber\\
&  +\int_{r}^{\infty}V\left(  r^{\prime}\right)  r^{\prime}\mathrm{d}%
r^{\prime}\int_{0}^{\infty}\mathrm{d}k^{2}\left[  \frac{\left(  kr\right)
^{n-2}\left(  kr^{\prime}\right)  ^{n-2}}{\pi\Gamma\left(  2\mu+1\right)
}\int_{0}^{\pi}\frac{Y_{\frac{n}{2}-1}\left(  qr^{\prime}\right)  }{\left(
qr^{\prime}\right)  ^{\frac{n}{2}-1}}\frac{J_{\frac{n}{2}-1}\left(  qr\right)
}{\left(  qr\right)  ^{\frac{n}{2}-1}}\sin^{n-3}\theta\mathrm{d}\cos
\theta\right]  \left.  \int_{0}^{\infty}\mathrm{e}^{-k^{2}t}t^{s}%
\mathrm{d}t\right\}  .
\end{align}
Performing the integral over $t$, we have%
\begin{align}
W^{\left(  2\right)  }\left(  s\right)   &  =\frac{\pi}{8}\tilde{\mu}%
^{2s}\Gamma\left(  s+1\right)  \int_{0}^{\infty}V\left(  r\right)
r\mathrm{d}r\nonumber\\
&  \times\left\{  \int_{0}^{r}V\left(  r^{\prime}\right)  r^{\prime}%
\mathrm{d}r^{\prime}\int_{0}^{\infty}\mathrm{d}k^{2}\left(  k^{2}\right)
^{-s-1}\left[  \frac{\left(  kr\right)  ^{n-2}\left(  kr^{\prime}\right)
^{n-2}}{\pi\Gamma\left(  2\mu+1\right)  }\int_{0}^{\pi}\frac{Y_{\frac{n}{2}%
-1}\left(  qr\right)  }{\left(  qr\right)  ^{\frac{n}{2}-1}}\frac{J_{\frac
{n}{2}-1}\left(  qr^{\prime}\right)  }{\left(  qr^{\prime}\right)  ^{\frac
{n}{2}-1}}\sin^{n-3}\theta\mathrm{d}\cos\theta\right]  \right. \nonumber\\
&  +\left.  \int_{r}^{\infty}V\left(  r^{\prime}\right)  r^{\prime}%
\mathrm{d}r^{\prime}\int_{0}^{\infty}\mathrm{d}k^{2}\left(  k^{2}\right)
^{-s-1}\left[  \frac{\left(  kr\right)  ^{n-2}\left(  kr^{\prime}\right)
^{n-2}}{\pi\Gamma\left(  2\mu+1\right)  }\int_{0}^{\pi}\frac{Y_{\frac{n}{2}%
-1}\left(  qr^{\prime}\right)  }{\left(  qr^{\prime}\right)  ^{\frac{n}{2}-1}%
}\frac{J_{\frac{n}{2}-1}\left(  qr\right)  }{\left(  qr\right)  ^{\frac{n}%
{2}-1}}\sin^{n-3}\theta\mathrm{d}\cos\theta\right]  \right\}  .
\end{align}

For odd-dimensional cases, substituting Eq. (\ref{ndkb2222}) into Eq.
(\ref{wsandrehe}) gives the second-order Born approximation of the one-loop
effective action:%
\begin{align}
W^{\left(  2\right)  }\left(  s\right)   &  =\frac{\tilde{\mu}^{2s}}%
{8\sqrt{\pi}}\frac{\Gamma\left(  \frac{n}{2}-\frac{1}{2}\right)  }%
{\Gamma\left(  n-1\right)  }\int_{0}^{\infty}\mathrm{d}rr^{\frac{n+1}{2}%
}V\left(  r\right) \nonumber\\
&  \times\left\{  \int_{0}^{r}\mathrm{d}r^{\prime}\left(  r^{\prime}\right)
^{\frac{n+1}{2}}V\left(  r^{\prime}\right)  \int_{0}^{\pi}P_{\frac{n}{2}%
-\frac{3}{2}}\left(  \cos\phi\right)  \mathrm{d}\cos\phi\right.  \int%
_{0}^{\infty}\mathrm{d}tt^{s-1}\left[  t^{1-\frac{n}{2}}\frac{\exp\left(
-\frac{1}{t}\left(  r^{2}+r^{\prime2}-2rr^{\prime}\cos\phi\right)  \right)
}{\sqrt{r^{2}+r^{\prime2}-2rr^{\prime}\cos\phi}}\right] \nonumber\\
&  +\int_{r}^{\infty}\mathrm{d}r^{\prime}\left(  r^{\prime}\right)
^{\frac{n+1}{2}}V\left(  r^{\prime}\right)  \int_{0}^{\pi}P_{\frac{n}{2}%
-\frac{3}{2}}\left(  \cos\phi\right)  \mathrm{d}\cos\phi\left.  \int%
_{0}^{\infty}\mathrm{d}tt^{s-1}\left[  t^{1-\frac{n}{2}}\frac{\exp\left(
-\frac{1}{t}\left(  r^{2}+r^{\prime2}-2rr^{\prime}\cos\phi\right)  \right)
}{\sqrt{r^{2}+r^{\prime2}-2rr^{\prime}\cos\phi}}\right]  \right\}  .
\end{align}
Performing the integral over $t$ gives%
\begin{align}
W^{\left(  2\right)  }\left(  s\right)   &  =\frac{\tilde{\mu}^{2s}}%
{8\sqrt{\pi}}\frac{\Gamma\left(  \frac{n}{2}-1-s\right)  \Gamma\left(
\frac{n}{2}-\frac{1}{2}\right)  }{\Gamma\left(  n-1\right)  }\int_{0}^{\infty
}\mathrm{d}rr^{\frac{n+1}{2}}V\left(  r\right) \nonumber\\
&  \times\left\{  \int_{0}^{r}\mathrm{d}r^{\prime}\left(  r^{\prime}\right)
^{\frac{n+1}{2}}V\left(  r^{\prime}\right)  \int_{0}^{\pi}\left(
r^{2}+r^{\prime2}-2rr^{\prime}\cos\phi\right)  ^{s-\left(  n-1\right)
/2}P_{\frac{n}{2}-\frac{3}{2}}\left(  \cos\phi\right)  \mathrm{d}\cos
\phi\right. \nonumber\\
&  \left.  +\int_{r}^{\infty}\mathrm{d}r^{\prime}\left(  r^{\prime}\right)
^{\frac{n+1}{2}}V\left(  r^{\prime}\right)  \int_{0}^{\pi}\left(
r^{2}+r^{\prime2}-2rr^{\prime}\cos\phi\right)  ^{s-\left(  n-1\right)
/2}P_{\frac{n}{2}-\frac{3}{2}}\left(  \cos\phi\right)  \mathrm{d}\cos
\phi\right\}  .
\end{align}

Around $s=0$, for $n\neq1$, we have
\begin{align}
W^{\left(  2\right)  }  &  =\frac{1}{8\sqrt{\pi}}\frac{\Gamma\left(  \frac
{n}{2}-1\right)  \Gamma\left(  \frac{n}{2}-\frac{1}{2}\right)  }{\Gamma\left(
n-1\right)  }\int_{0}^{\infty}\mathrm{d}rr^{\frac{n+1}{2}}V\left(  r\right)
\left\{  \int_{0}^{r}\mathrm{d}r^{\prime}\left(  r^{\prime}\right)
^{\frac{n+1}{2}}V\left(  r^{\prime}\right)  \int_{0}^{\pi}\frac{P_{\frac{n}%
{2}-\frac{3}{2}}\left(  \cos\phi\right)  }{\left(  r^{2}+r^{\prime
2}-2rr^{\prime}\cos\phi\right)  ^{\left(  n-1\right)  /2}}\mathrm{d}\cos
\phi\right. \nonumber\\
&  \left.  +\int_{r}^{\infty}\mathrm{d}r^{\prime}\left(  r^{\prime}\right)
^{\frac{n+1}{2}}V\left(  r^{\prime}\right)  \int_{0}^{\pi}\frac{P_{\frac{n}%
{2}-\frac{3}{2}}\left(  \cos\phi\right)  }{\left(  r^{2}+r^{\prime
2}-2rr^{\prime}\cos\phi\right)  ^{\left(  n-1\right)  /2}}\mathrm{d}\cos
\phi\right\}  .
\end{align}

Next, we perform the angle integral. Rewrite the angle integral as
\begin{equation}
\int_{0}^{\pi}\frac{P_{\frac{n}{2}-\frac{3}{2}}\left(  \cos\phi\right)
}{\left(  r^{2}+r^{\prime2}-2rr^{\prime}\cos\phi\right)  ^{\frac{n}{2}%
-\frac{1}{2}}}\mathrm{d}\cos\phi=\frac{1}{\left(  2rr^{\prime}\right)
^{\frac{n}{2}-\frac{1}{2}}}\int_{0}^{\pi}\frac{P_{\frac{n}{2}-\frac{3}{2}%
}\left(  \cos\phi\right)  }{\left(  R-\cos\phi\right)  ^{\frac{n}{2}-\frac
{1}{2}}}\mathrm{d}\cos\phi, \label{lpandR}%
\end{equation}
where $R=\frac{r^{2}+r^{\prime2}}{2rr^{\prime}}$. It can be checked that%
\begin{align}
\frac{1}{\left(  R-\cos\phi\right)  ^{\frac{n}{2}-\frac{1}{2}}}  &  =\frac
{1}{\left(  -1\right)  ^{\left(  \frac{n}{2}-\frac{3}{2}\right)  }}%
\frac{\mathrm{d}^{\left(  n-3\right)  /2}}{\mathrm{d}R^{\left(  n-3\right)
/2}}\frac{1}{R-\cos\phi},\nonumber\\
\left(  -1\right)  ^{\left(  \frac{n}{2}-\frac{3}{2}\right)  }  &  =\left(
-1\right)  \times\left(  -2\right)  \times\left(  -3\right)  \times
...\times\left(  -1-\left(  \frac{n}{2}-\frac{3}{2}\right)  \right)  ,
\end{align}
so the integral (\ref{lpandR}) becomes%
\begin{align}
\int_{0}^{\pi}\frac{P_{\frac{n}{2}-\frac{3}{2}}\left(  \cos\phi\right)
}{\left(  r^{2}+r^{\prime2}-2rr^{\prime}\cos\phi\right)  ^{\frac{n}{2}%
-\frac{1}{2}}}\mathrm{d}\cos\phi &  =\frac{1}{\left(  -1\right)  ^{\left(
\frac{n}{2}-\frac{3}{2}\right)  }\left(  2rr^{\prime}\right)  ^{\frac{n}%
{2}-\frac{1}{2}}}\frac{\mathrm{d}^{\left(  n-3\right)  /2}}{\mathrm{d}%
R^{\left(  n-3\right)  /2}}\int_{0}^{\pi}\frac{P_{\frac{n}{2}-\frac{3}{2}%
}\left(  \cos\phi\right)  }{R-\cos\phi}\mathrm{d}\cos\phi\nonumber\\
&  =\frac{1}{2^{\frac{n}{2}-\frac{3}{2}}\left(  -1\right)  ^{\left(  \frac
{n}{2}-\frac{3}{2}\right)  }\left(  rr^{\prime}\right)  ^{\frac{n}{2}-\frac
{1}{2}}}\frac{\mathrm{d}^{\left(  n-3\right)  /2}}{\mathrm{d}R^{\left(
n-3\right)  /2}}Q_{\frac{n}{2}-\frac{3}{2}}\left(  R\right)  ,
\end{align}
where $Q_{\nu}\left(  z\right)  $ is the Legendre function of the second kind:
$\mathbb{P}\int_{0}^{\pi}\frac{1}{R-\cos\phi}P_{\frac{n}{2}-\frac{3}{2}%
}\left(  \cos\phi\right)  \mathrm{d}\cos\phi=2Q_{\frac{n}{2}-\frac{3}{2}%
}\left(  R\right)  $ \cite{olver2010nist}. Then the second-order one-loop
effective action reads%
\begin{align}
W^{\left(  2\right)  }  &  \sim\frac{1}{2^{\frac{n}{2}+\frac{3}{2}}\left(
-1\right)  ^{\left(  \frac{n}{2}-\frac{3}{2}\right)  }\sqrt{\pi}}\frac
{\Gamma\left(  \frac{n}{2}-1\right)  \Gamma\left(  \frac{n}{2}-\frac{1}%
{2}\right)  }{\Gamma\left(  n-1\right)  }\nonumber\\
&  \times\int_{0}^{\infty}r\mathrm{d}rV\left(  r\right)  \left\{  \int_{0}%
^{r}r^{\prime}\mathrm{d}r^{\prime}V\left(  r^{\prime}\right)  \frac
{\mathrm{d}^{m-1}}{\mathrm{d}R^{m-1}}Q_{m-1}\left(  R\right)  +\int%
_{r}^{\infty}r^{\prime}\mathrm{d}r^{\prime}V\left(  r^{\prime}\right)
\frac{\mathrm{d}^{m-1}}{\mathrm{d}R^{m-1}}Q_{m-1}\left(  R\right)  \right\}  .
\end{align}

\subsubsection{Vacuum energy}

Similarly, using the relation between the heat kernel and the vacuum energy,
Eq. (\ref{eandrehe}), we can obtain the second-order $n$-dimensional Born
approximation of the vacuum energy. Substituting Eq. (\ref{ndkb2}) into Eq.
(\ref{eandrehe}) gives%
\begin{align}
E_{0}^{\left(  2\right)  }\left(  \epsilon\right)   &  =-\frac{\pi}{8}%
\tilde{\mu}^{2\epsilon}\frac{1}{\Gamma\left(  -\frac{1}{2}+\epsilon\right)
}\int_{0}^{\infty}V\left(  r\right)  r\mathrm{d}r\nonumber\\
&  \times\left\{  \int_{0}^{r}V\left(  r^{\prime}\right)  r^{\prime}%
\mathrm{d}r^{\prime}\int_{0}^{\infty}\mathrm{d}k^{2}\left[  \frac{\left(
kr\right)  ^{n-2}\left(  kr^{\prime}\right)  ^{n-2}}{\pi\Gamma\left(
2\mu+1\right)  }\int_{0}^{\pi}\frac{Y_{\frac{n}{2}-1}\left(  qr\right)
}{\left(  qr\right)  ^{\frac{n}{2}-1}}\frac{J_{\frac{n}{2}-1}\left(
qr^{\prime}\right)  }{\left(  qr^{\prime}\right)  ^{\frac{n}{2}-1}}\sin
^{n-3}\theta\mathrm{d}\cos\theta\right]  \right.  \int_{0}^{\infty}%
\mathrm{e}^{-k^{2}t}t^{-\frac{1}{2}+\epsilon}\mathrm{d}t\nonumber\\
&  +\int_{r}^{\infty}V\left(  r^{\prime}\right)  r^{\prime}\mathrm{d}%
r^{\prime}\int_{0}^{\infty}\mathrm{d}k^{2}\left[  \frac{\left(  kr\right)
^{n-2}\left(  kr^{\prime}\right)  ^{n-2}}{\pi\Gamma\left(  2\mu+1\right)
}\int_{0}^{\pi}\frac{Y_{\frac{n}{2}-1}\left(  qr\right)  }{\left(  qr\right)
^{\frac{n}{2}-1}}\frac{J_{\frac{n}{2}-1}\left(  qr^{\prime}\right)  }{\left(
qr^{\prime}\right)  ^{\frac{n}{2}-1}}\sin^{n-3}\theta\mathrm{d}\cos
\theta\right]  \left.  \int_{0}^{\infty}\mathrm{e}^{-k^{2}t}t^{-\frac{1}%
{2}+\epsilon}\mathrm{d}t\right\}  .
\end{align}
Performing the integral, we have%
\begin{align}
E_{0}^{\left(  2\right)  }\left(  \epsilon\right)   &  =-\frac{\pi}{8}%
\tilde{\mu}^{2\epsilon}\frac{\Gamma\left(  \frac{1}{2}+\epsilon\right)
}{\Gamma\left(  -\frac{1}{2}+\epsilon\right)  }\int_{0}^{\infty}V\left(
r\right)  r\mathrm{d}r\nonumber\\
&  \times\left\{  \int_{0}^{r}V\left(  r^{\prime}\right)  r^{\prime}%
\mathrm{d}r^{\prime}\int_{0}^{\infty}\mathrm{d}k^{2}\left(  k^{2}\right)
^{-\frac{1}{2}-\epsilon}\left[  \frac{\left(  kr\right)  ^{n-2}\left(
kr^{\prime}\right)  ^{n-2}}{\pi\Gamma\left(  2\mu+1\right)  }\int_{0}^{\pi
}\frac{Y_{\frac{n}{2}-1}\left(  qr\right)  }{\left(  qr\right)  ^{\frac{n}%
{2}-1}}\frac{J_{\frac{n}{2}-1}\left(  qr^{\prime}\right)  }{\left(
qr^{\prime}\right)  ^{\frac{n}{2}-1}}\sin^{n-3}\theta\mathrm{d}\cos
\theta\right]  \right. \nonumber\\
&  \left.  +\int_{r}^{\infty}V\left(  r^{\prime}\right)  r^{\prime}%
\mathrm{d}r^{\prime}\int_{0}^{\infty}\mathrm{d}k^{2}\left(  k^{2}\right)
^{-\frac{1}{2}-\epsilon}\left[  \frac{\left(  kr\right)  ^{n-2}\left(
kr^{\prime}\right)  ^{n-2}}{\pi\Gamma\left(  2\mu+1\right)  }\int_{0}^{\pi
}\frac{Y_{\frac{n}{2}-1}\left(  qr\right)  }{\left(  qr\right)  ^{\frac{n}%
{2}-1}}\frac{J_{\frac{n}{2}-1}\left(  qr^{\prime}\right)  }{\left(
qr^{\prime}\right)  ^{\frac{n}{2}-1}}\sin^{n-3}\theta\mathrm{d}\cos
\theta\right]  \right\}  .
\end{align}

For odd-dimensional cases, substituting Eq. (\ref{ndkb2222}) into Eq.
(\ref{eandrehe}) gives the odd-dimensional second-order Born approximation of
the vacuum energy:%
\begin{align}
E_{0}^{\left(  2\right)  }\left(  \epsilon\right)   &  =-\frac{\tilde{\mu
}^{2\epsilon}}{8\sqrt{\pi}}\frac{\Gamma\left(  \frac{n}{2}-\frac{1}{2}\right)
}{\Gamma\left(  -\frac{1}{2}+\epsilon\right)  \Gamma\left(  n-1\right)  }%
\int_{0}^{\infty}r^{\frac{n+1}{2}}V\left(  r\right)  \mathrm{d}r\nonumber\\
&  \times\left\{  \int_{0}^{r}\left(  r^{\prime}\right)  ^{\frac{n+1}{2}%
}V\left(  r^{\prime}\right)  \mathrm{d}r^{\prime}\int_{0}^{\pi}P_{\frac{n}%
{2}-\frac{3}{2}}\left(  \cos\phi\right)  \mathrm{d}\cos\phi\right.  \int%
_{0}^{\infty}t^{-\frac{1}{2}+\epsilon-1}\left[  t^{1-\frac{n}{2}}\frac
{\exp\left(  -\frac{1}{t}\left(  r^{2}+r^{\prime2}-2rr^{\prime}\cos
\phi\right)  \right)  }{\sqrt{r^{2}+r^{\prime2}-2rr^{\prime}\cos\phi}}\right]
\mathrm{d}t\nonumber\\
&  +\int_{r}^{\infty}\left(  r^{\prime}\right)  ^{\frac{n+1}{2}}V\left(
r^{\prime}\right)  \mathrm{d}r^{\prime}\int_{0}^{\pi}P_{\frac{n}{2}-\frac
{3}{2}}\left(  \cos\phi\right)  \mathrm{d}\cos\phi\left.  \int_{0}^{\infty
}t^{-\frac{1}{2}+\epsilon-1}\left[  t^{1-\frac{n}{2}}\frac{\exp\left(
-\frac{1}{t}\left(  r^{2}+r^{\prime2}-2rr^{\prime}\cos\phi\right)  \right)
}{\sqrt{r^{2}+r^{\prime2}-2rr^{\prime}\cos\phi}}\right]  \mathrm{d}t\right\}
.
\end{align}
Performing the integral over $t$ gives%
\begin{align}
E_{0}^{\left(  2\right)  }\left(  \epsilon\right)   &  =-\frac{\tilde{\mu
}^{2\epsilon}}{8\sqrt{\pi}}\frac{\Gamma\left(  \frac{n}{2}-\frac{1}%
{2}-\epsilon\right)  \Gamma\left(  \frac{n}{2}-\frac{1}{2}\right)  }%
{\Gamma\left(  \epsilon-\frac{1}{2}\right)  \Gamma\left(  n-1\right)  }%
\int_{0}^{\infty}r^{\frac{n+1}{2}}V\left(  r\right)  \mathrm{d}r\nonumber\\
&  \times\left\{  \int_{0}^{r}\left(  r^{\prime}\right)  ^{\frac{n+1}{2}%
}V\left(  r^{\prime}\right)  \mathrm{d}r^{\prime}\int_{0}^{\pi}\left(
r^{2}+r^{\prime2}-2rr^{\prime}\cos\phi\right)  ^{-\frac{n}{2}+\epsilon
}P_{\frac{n}{2}-\frac{3}{2}}\left(  \cos\phi\right)  \mathrm{d}\cos\phi\right.
\nonumber\\
&  \left.  +\int_{r}^{\infty}\left(  r^{\prime}\right)  ^{\frac{n+1}{2}%
}V\left(  r^{\prime}\right)  \mathrm{d}r^{\prime}\int_{0}^{\pi}\left(
r^{2}+r^{\prime2}-2rr^{\prime}\cos\phi\right)  ^{-\frac{n}{2}+\epsilon
}P_{\frac{n}{2}-\frac{3}{2}}\left(  \cos\phi\right)  \mathrm{d}\cos
\phi\right\}  .
\end{align}

Around $\epsilon=-\frac{1}{2}$, for $n\neq1$, we have%
\begin{align}
E_{0}^{\left(  2\right)  }  &  =\frac{1}{2^{\frac{n+5}{2}}\sqrt{\pi}\tilde
{\mu}}\frac{\Gamma\left(  \frac{n}{2}-\frac{1}{2}\right)  }{\Gamma\left(
n-1\right)  }\int_{0}^{\infty}\mathrm{d}rr^{\frac{n+1}{2}}V\left(  r\right)
\left\{  \int_{0}^{r}\mathrm{d}r^{\prime}\left(  r^{\prime}\right)
^{\frac{n+1}{2}}V\left(  r^{\prime}\right)  \int_{0}^{\pi}\frac{P_{\frac{n}%
{2}-\frac{3}{2}}\left(  \cos\phi\right)  }{\left(  r^{2}+r^{\prime
2}-2rr^{\prime}\cos\phi\right)  ^{\frac{n}{2}+\frac{1}{2}}}\mathrm{d}\cos
\phi\right. \nonumber\\
&  \left.  +\int_{r}^{\infty}\mathrm{d}r^{\prime}\left(  r^{\prime}\right)
^{\frac{n+1}{2}}V\left(  r^{\prime}\right)  \int_{0}^{\pi}\frac{P_{\frac{n}%
{2}-\frac{3}{2}}\left(  \cos\phi\right)  }{\left(  r^{2}+r^{\prime
2}-2rr^{\prime}\cos\phi\right)  ^{\frac{n}{2}+\frac{1}{2}}}\mathrm{d}\cos
\phi\right\}  .
\end{align}
A similar treatment gives
\begin{align}
\int_{0}^{\pi}\frac{P_{\frac{n}{2}-\frac{3}{2}}\left(  \cos\phi\right)
}{\left(  r^{2}+r^{\prime2}-2rr^{\prime}\cos\phi\right)  ^{\frac{n}{2}%
+\frac{1}{2}}}\mathrm{d}\cos\phi &  =\frac{1}{2^{\frac{n}{2}+\frac{1}{2}%
}\left(  -1\right)  ^{\left(  \frac{n}{2}-\frac{1}{2}\right)  }\left(
rr^{\prime}\right)  ^{\frac{n}{2}+\frac{1}{2}}}\frac{\mathrm{d}^{\left(
n-1\right)  /2}}{\mathrm{d}R^{\left(  n-1\right)  /2}}\int_{0}^{\pi}%
\frac{P_{\frac{n}{2}-\frac{3}{2}}\left(  \cos\phi\right)  }{R-\cos\phi
}\mathrm{d}\cos\phi\nonumber\\
&  =\frac{1}{2^{\frac{n}{2}-\frac{1}{2}}\left(  -1\right)  ^{\left(  \frac
{n}{2}-\frac{1}{2}\right)  }\left(  rr^{\prime}\right)  ^{\frac{n}{2}+\frac
{1}{2}}}\frac{\mathrm{d}^{\left(  n-1\right)  /2}}{\mathrm{d}R^{\left(
n-1\right)  /2}}Q_{\frac{n}{2}-\frac{3}{2}}\left(  R\right)  .
\end{align}
Then the second-order vacuum energy reads%
\begin{align}
E_{0}^{\left(  2\right)  }  &  \sim\frac{1}{2^{n+2}\left(  -1\right)
^{\left(  \frac{n}{2}-\frac{1}{2}\right)  }\sqrt{\pi}\tilde{\mu}}\frac
{\Gamma\left(  \frac{n}{2}-\frac{1}{2}\right)  }{\Gamma\left(  n-1\right)
}\nonumber\\
&  \times\int_{0}^{\infty}r\mathrm{d}rV\left(  r\right)  \left\{  \int_{0}%
^{r}\mathrm{d}r^{\prime}V\left(  r^{\prime}\right)  \frac{\mathrm{d}^{\left(
n-1\right)  /2}}{\mathrm{d}R^{\left(  n-1\right)  /2}}Q_{m-1}\left(  R\right)
+\int_{r}^{\infty}\mathrm{d}r^{\prime}V\left(  r^{\prime}\right)
\frac{\mathrm{d}^{\left(  n-1\right)  /2}}{\mathrm{d}R^{\left(  n-1\right)
/2}}Q_{m-1}\left(  R\right)  \right\}  .
\end{align}

\section{WKB approximation \label{WKB}}

In this section, we convert the WKB approximation method into a method for
calculating one-loop effective actions and vacuum energies.

The WKB approximation of the scattering phase shift is \cite{landau1965course}%
\begin{equation}
\delta_{l}^{\text{WKB}}\left(  k\right)  \sim-\frac{1}{2k}\int_{\left(
l+1/2\right)  /k}^{\infty}\frac{rV\left(  r\right)  }{\sqrt{r^{2}-\left(
\frac{l+1/2}{k}\right)  ^{2}}}\mathrm{d}r. \label{pswkb}%
\end{equation}
Substituting the phase shift (\ref{pswkb}) into Eq. (\ref{psdaorehe}) gives
the WKB approximation of the heat kernel:%
\begin{equation}
K_{l}^{\text{WKB}}\left(  t\right)  \sim-\frac{t}{2\pi}\int_{0}^{\infty
}\mathrm{d}k^{2}\frac{1}{k}\mathrm{e}^{-k^{2}t}\sum_{l=0}^{\infty}\left(
2l+1\right)  \int_{\left(  l+1/2\right)  /k}^{\infty}\frac{rV\left(  r\right)
}{\sqrt{r^{2}-\left(  \frac{l+1/2}{k}\right)  ^{2}}}\mathrm{d}r. \label{KlWKB}%
\end{equation}
Substituting the phase shift (\ref{pswkb}) into Eq. (\ref{1loop}) gives the
WKB approximation of the one-loop effective action:%
\begin{equation}
W_{s}^{\text{WKB}}\sim\frac{1}{4\pi}\tilde{\mu}^{2s}\Gamma\left(  s+1\right)
\int_{0}^{\infty}\mathrm{d}k^{2}\frac{1}{\left(  k^{2}\right)  ^{s+3/2}}%
\sum_{l=0}^{\infty}\left(  2l+1\right)  \int_{\left(  l+1/2\right)
/k}^{\infty}\frac{rV\left(  r\right)  }{\sqrt{r^{2}-\left(  \frac{l+1/2}%
{k}\right)  ^{2}}}\mathrm{d}r. \label{olewkb}%
\end{equation}
Substituting the phase shift (\ref{pswkb}) into Eq. (\ref{vE}) gives the WKB
approximation of the vacuum energy:
\begin{equation}
E_{0}^{\text{WKB}}\left(  \epsilon\right)  \sim-\frac{1}{4\pi}\tilde{\mu
}^{2\epsilon}\frac{\Gamma\left(  \frac{1}{2}+\epsilon\right)  }{\Gamma\left(
-\frac{1}{2}+\epsilon\right)  }\int_{0}^{\infty}\mathrm{d}k^{2}\frac
{1}{\left(  k^{2}\right)  ^{1+\epsilon}}\sum_{l=0}^{\infty}\left(
2l+1\right)  \int_{\left(  l+1/2\right)  /k}^{\infty}\frac{rV\left(  r\right)
}{\sqrt{r^{2}-\left(  \frac{l+1/2}{k}\right)  ^{2}}}\mathrm{d}r. \label{vewkb}%
\end{equation}

\subsection{Example: $V\left(  r\right)  =\frac{\alpha}{\left(  r^{2}%
+r_{0}^{2}\right)  ^{2}}$}

In this section, we consider the potential
\begin{equation}
V\left(  r\right)  =\frac{\alpha}{\left(  r^{2}+r_{0}^{2}\right)  ^{2}}
\label{pot}%
\end{equation}
as an example.

Substituting the potential (\ref{pot}) into Eq. (\ref{KlWKB}) gives%
\begin{align}
K_{l}^{\text{WKB}}\left(  t\right)   &  \sim-\frac{t}{2\pi}\int_{0}^{\infty
}\mathrm{d}k^{2}\frac{1}{k}\mathrm{e}^{-k^{2}t}\sum_{l=0}^{\infty}\left(
2l+1\right)  \frac{\pi\alpha}{4\left[  r_{0}^{2}+\left(  \frac{l+1/2}%
{k}\right)  ^{2}\right]  ^{3/2}}\nonumber\\
&  =-\frac{t}{2\pi}\int_{0}^{\infty}\frac{\alpha k^{2}\pi}{\sqrt{1+4r_{0}%
^{2}k^{2}}}\mathrm{e}^{-k^{2}t}\mathrm{d}k^{2}\nonumber\\
&  =-\frac{\alpha}{8r_{0}^{2}}+\frac{\alpha\sqrt{\pi}\mathrm{e}^{\frac
{t}{4r_{0}^{2}}}\operatorname*{erfc}\left(  \frac{\sqrt{t}}{2r_{0}}\right)
\sqrt{t}}{16r_{0}^{3}}-\frac{\alpha\sqrt{\pi}\mathrm{e}^{\frac{t}{4r_{0}^{2}}%
}\operatorname*{erfc}\left(  \frac{\sqrt{t}}{2r_{0}}\right)  }{8r_{0}\sqrt{t}%
}, \label{ktWKB}%
\end{align}
where $\operatorname*{erfc}\left(  z\right)  $ is the complementary error
function \cite{olver2010nist}.

Similarly, by Eqs. (\ref{olewkb}) and (\ref{vewkb}), we can obtain the
one-loop effective action and the vacuum energy:%
\begin{equation}
W_{s}^{\text{WKB}}\sim\frac{\alpha\tilde{\mu}^{2s}r_{0}^{2s-2}}{4}%
s\Gamma\left(  1-s\right)  \Gamma\left(  2s-1\right)
\end{equation}
and%
\begin{equation}
E_{0}^{\text{WKB}}\left(  \epsilon\right)  \sim-\frac{4^{\epsilon-3}%
\alpha\tilde{\mu}^{2\epsilon}r_{0}^{2\epsilon-3}}{\sqrt{\pi}}\left(
2\epsilon-1\right)  \Gamma\left(  \frac{3}{2}-\epsilon\right)  \Gamma\left(
\epsilon-1\right)  .
\end{equation}

\subsection{WKB approximation and Born approximation: comparison}

Taking the heat kernel as an example, we compare the WKB approximation with
the Born approximation.

The heat kernel given by the first-order Born approximation can be obtained by
substituting the potential (\ref{pot}) into Eq. (\ref{ktborn1}):%
\begin{equation}
K^{\text{s}\left(  1\right)  }\left(  t\right)  \sim-\frac{\alpha\sqrt{\pi}%
}{8r_{0}\sqrt{t}}.
\end{equation}
\ 

Expanding the heat kernel obtained by the WKB approximation, Eq.
(\ref{ktWKB}), we have%
\begin{equation}
K_{l}^{\text{WKB}}\left(  t\right)  \sim-\frac{\alpha\sqrt{\pi}}{8r_{0}%
\sqrt{t}}+\frac{\alpha\sqrt{\pi}}{32r_{0}^{3}}\sqrt{t}%
\end{equation}
The leading order contribution of these two methods are the same.

It should be noted that both WKB approximation and Born approximation are
valid for short-range potentials. The short-range potential at $r\rightarrow0$
satisfies \cite{chadan2012inverse,li2021long}%
\begin{equation}
\int_{0}^{a}\left\vert V\left(  r\right)  \right\vert r\mathrm{d}%
r<\infty,\text{ }a<\infty,
\end{equation}
and at $r\rightarrow\infty$ satisfies%
\begin{equation}
\int_{b}^{\infty}\left\vert V\left(  r\right)  \right\vert \mathrm{d}%
r<\infty,\text{ }b>0.
\end{equation}

For long-range potentials, divergence may be encountered, and the
renormalization procedure is required to remove the divergence. The treatment
of removing divergence can refer to Ref. \cite{li2016scattering}.

\section{One-loop effective action and vacuum energy of scalar field in curved
spacetime}

\subsection{Schwarzschild spacetime}

Ref. \cite{li2018scalar} gives a first-order scattering phase shift of a
scalar field on a Schwarzschild black hole:
\begin{equation}
\delta_{l}^{(1)}=-\arctan\left(  \frac{\int_{1}^{\infty}\sin^{2}\left(
2M\eta\left[  \rho+\ln\left(  \rho-1\right)  \right]  \right)  \frac{1}%
{\rho-1}V_{l}^{\text{eff}}\left(  \rho\right)  \rho\mathrm{d}\rho}%
{2M\eta+\frac{1}{2}\int_{1}^{\infty}\sin\left(  4M\eta\left[  \rho+\ln\left(
\rho-1\right)  \right]  \right)  \frac{1}{\rho-1}V_{l}^{\text{eff}}\left(
\rho\right)  \rho\mathrm{d}\rho}\right)  ,
\end{equation}
where $V_{l}^{\text{eff}}\left(  \rho\right)  =\left(  1-\frac{1}{\rho
}\right)  \left[  \frac{l\left(  l+1\right)  }{\rho^{2}}+\frac{1}{\rho^{3}%
}\right]  -\frac{\left(  2M\mu\right)  ^{2}}{\rho}$ with $\rho=r/2M$, $M$ the
mass of the black hole, and $\mu$ the mass of the scalar field.

Consider a massless scalar field ($\mu=0$) in a Schwarzschild spacetime with
$l=0$, we have $\eta=\sqrt{k^{2}-\mu^{2}}=k$. Then%
\begin{equation}
\delta_{0}^{(1)}=-\frac{1}{2Mk}\int_{1}^{\infty}\sin^{2}\left(  2Mk\left[
\rho+\ln\left(  \rho-1\right)  \right]  \right)  \frac{1}{\rho^{3}}%
\mathrm{d}\rho. \label{delta1}%
\end{equation}
Substituting Eq. (\ref{delta1}) into Eq. (\ref{hkbyps}) and performing the
integral over $k$ give the heat kernel,%
\begin{equation}
K_{0}^{s}\left(  t\right)  =\frac{\sqrt{t}}{4M\sqrt{\pi}}\left(  \int%
_{1}^{\infty}\frac{1}{\rho^{3}}\mathrm{e}^{-\frac{4M^{2}}{t}\left(  \rho
+\ln\left(  \rho-1\right)  \right)  ^{2}}\mathrm{d}\rho-\frac{1}{2}\right)  .
\end{equation}

The one-loop effective action for $l=0$ by Eq. (\ref{wsandrehe}) reads%
\begin{equation}
W_{0}\left(  s\right)  =-\widetilde{\mu}^{2s}\frac{2^{2\left(  s-1\right)
}M^{2s}}{\sqrt{\pi}}\Gamma\left(  -s-\frac{1}{2}\right)  \int_{1}^{\infty
}\frac{1}{\rho^{3}}\left[  \rho+\ln\left(  \rho-1\right)  \right]
^{2s+1}\mathrm{d}\rho.
\end{equation}
The regularized vacuum energy for $l=0$ by Eq. (\ref{eandrehe}) reads%
\begin{equation}
E_{0}\left(  \epsilon\right)  =\widetilde{\mu}^{2\epsilon}\frac{\Gamma\left(
-\epsilon\right)  }{\Gamma\left(  -\frac{1}{2}+\epsilon\right)  }%
\frac{2^{2\epsilon-3}}{\sqrt{\pi}}M^{2\epsilon-1}\int_{1}^{\infty}\frac
{1}{\rho^{3}}\left[  \rho+\ln\left(  \rho-1\right)  \right]  ^{2\epsilon
}\mathrm{d}\rho.
\end{equation}

\subsection{Reissner-Nordstr\"{o}m spacetime}

Ref. \cite{li2021scalar} gives a first-order scattering phase shift of a
scalar field on a Reissner-Nordstr\"{o}m black hole:%
\begin{equation}
\delta_{l}^{(1)}=-\arctan\frac{\frac{1}{\eta}\int_{r_{+}}^{\infty}\sin
^{2}\left(  \eta r_{\ast}\right)  \frac{\mathrm{d}r_{\ast}}{\mathrm{d}r}%
V_{l}^{\text{eff}}\mathrm{d}r}{1+\frac{1}{\eta}\int_{r_{+}}^{\infty}%
\sin\left(  2\eta r_{\ast}\right)  \frac{\mathrm{d}r_{\ast}}{\mathrm{d}r}%
V_{l}^{\text{eff}}\mathrm{d}r}+\left(  r_{+}+r_{-}\right)  \eta\ln\frac
{r_{+}-r_{-}}{r_{+}+r_{-}},
\end{equation}
where $V_{l}^{\text{eff}}=\left(  1-\frac{r_{+}}{r}\right)  \left(
1-\frac{r_{-}}{r}\right)  \left[  \frac{l\left(  l+1\right)  }{r^{2}}+\left(
\frac{r_{+}+r_{-}}{r^{3}}-\frac{2r_{+}r_{-}}{r^{4}}\right)  \right]  $,
$r_{\pm}=M\pm\sqrt{M^{2}-Q^{2}}$ with $M$ the mass and $Q$ the charge of the
spacetime, and the tortoise coordinate $r_{\ast}=r+\frac{r_{+}^{2}}%
{r_{+}-r_{-}}\ln\left(  \frac{r}{r_{+}}-1\right)  -\frac{r_{-}^{2}}%
{r_{+}-r_{-}}\ln\left(  \frac{r}{r_{-}}-1\right)  $.

Consider a massless scalar field ($\mu=0$) in a Reissner-Nordstr\"{o}m
spacetime with $l=0$, we have $\eta=\sqrt{k^{2}-\mu^{2}}=k$. Then%
\begin{equation}
\delta_{0}^{(1)}=-\frac{1}{k}\int_{r_{+}}^{\infty}\sin^{2}\left(  kr_{\ast
}\right)  \left(  \frac{2M}{r^{3}}-\frac{2Q_{-}^{2}}{r^{4}}\right)
\mathrm{d}r+Mk\ln\left(  1-\frac{Q^{2}}{M^{2}}\right)  .
\end{equation}
Substituting into Eq.(\ref{hkbyps}) and performing the integral over $k$ give
the heat kernel,%
\[
K_{0}^{s}\left(  t\right)  =-\frac{\sqrt{t}}{2\sqrt{\pi}}\int_{r_{+}}^{\infty
}\mathrm{e}^{-\frac{r_{\ast}^{2}}{t}}\left(  \mathrm{e}^{\frac{r_{\ast}^{2}%
}{t}}-1\right)  \left(  \frac{2M}{r^{3}}-\frac{2Q^{2}}{r^{4}}\right)
\mathrm{d}r-\frac{M}{2\sqrt{\pi t}}\ln\left(  1-\frac{Q^{2}}{M^{2}}\right)  .
\]

The one-loop effective action for $l=0$ by Eq. (\ref{wsandrehe}) reads%
\begin{equation}
W_{0}\left(  s\right)  =-\frac{\widetilde{\mu}^{2s}}{4\sqrt{\pi}}\Gamma\left(
-s-\frac{1}{2}\right)  \int_{r_{+}}^{\infty}r_{\ast}^{2s+1}\left(  \frac
{2M}{r^{3}}-\frac{2Q^{2}}{r^{4}}\right)  \mathrm{d}r.
\end{equation}

The regularized vacuum energy for $l=0$ by Eq. (\ref{eandrehe}) reads%
\begin{equation}
E_{0}\left(  \epsilon\right)  =-\frac{\widetilde{\mu}^{2\epsilon}}{4\sqrt{\pi
}}\frac{1}{\Gamma\left(  -\frac{1}{2}+\epsilon\right)  }\int_{r_{+}}^{\infty
}r_{\ast}^{2\epsilon}\left(  \frac{2M}{r^{3}}-\frac{2Q^{2}}{r^{4}}\right)
\mathrm{d}r.
\end{equation}

\section{Calculating global heat kernel, one-loop effective action, and vacuum
energy from scattering amplitude \label{HOVPA}}

In the above, we calculate the heat kernel, the vacuum energy, and the
one-loop effective action from the scattering phase shift. In this section, we
suggest a method that calculates them from the scattering amplitude.

The scattering wave function is \cite{li2016scattering}%
\begin{equation}
\psi\left(  r,\theta\right)  =\mathrm{e}^{\mathrm{i}kr\cos\theta}+\sum
_{l=0}^{\infty}a_{l}\left(  \theta\right)  h_{l}^{\left(  1\right)  }\left(
kr\right)  , \label{sbcal}%
\end{equation}
where%
\begin{equation}
a_{l}\left(  \theta\right)  =\left(  2l+1\right)  \mathrm{i}^{l}\frac{1}%
{2}\left(  \mathrm{e}^{2\mathrm{i}\delta_{l}}-1\right)  P_{l}\left(
\cos\theta\right)  \label{altheta}%
\end{equation}
is the partial scattering amplitude.

Under the large-distance approximation, by the asymptotics of the Hankel
function $h_{l}^{\left(  1\right)  }\left(  kr\right)  $, Eq. (\ref{h12inf}),
we have%
\begin{equation}
\psi\left(  r,\theta\right)  =\mathrm{e}^{\mathrm{i}kr\cos\theta}+f\left(
\theta\right)  \frac{\mathrm{e}^{\mathrm{i}kr}}{r},
\end{equation}
where%
\begin{equation}
f\left(  \theta\right)  =\frac{1}{2\mathrm{i}k}\sum_{l=0}^{\infty}\left(
2l+1\right)  P_{l}\left(  \cos\theta\right)  \left(  \mathrm{e}^{2\mathrm{i}%
\delta_{l}}-1\right)  \label{sanshezhenfu}%
\end{equation}
is the scattering amplitude, for the differential scattering cross section
$\sigma\left(  \theta\right)  =\left\vert f\left(  \theta\right)  \right\vert
^{2}$.

For small phase shifts, we approximate $\mathrm{e}^{2\mathrm{i}\delta_{l}%
}\simeq1+2\mathrm{i}\delta_{l}$ in Eq. (\ref{sanshezhenfu}):%
\begin{equation}
f\left(  \theta\right)  \simeq\frac{1}{k}\sum_{l=0}^{\infty}\left(
2l+1\right)  \delta_{l}P_{l}\left(  \cos\theta\right)  .
\end{equation}
Then the forward-scattering amplitude, the scattering amplitude in the
direction $\theta=0$, is%

\begin{equation}
f\left(  0\right)  =\frac{1}{k}\sum_{l=0}^{\infty}\left(  2l+1\right)
\delta_{l}.
\end{equation}
Noting that $D_{l}=2l+1$ is the degeneracy, we have%
\begin{equation}
\sum_{l=0}^{\infty}\left(  2l+1\right)  \delta_{l}=kf\left(  0\right)  .
\end{equation}
Substituting into Eq. (\ref{psdaorehe}) gives%
\begin{align}
K\left(  t\right)   &  =\frac{t}{\pi}\int_{0}^{\infty}\mathrm{d}%
k^{2}\mathrm{e}^{-k^{2}t}\sum_{l=0}^{\infty}\left(  2l+1\right)  \delta
_{l}\nonumber\\
&  =\frac{2}{\pi}t\int_{0}^{\infty}f\left(  0\right)  \mathrm{e}^{-k^{2}%
t}k^{2}\mathrm{d}k. \label{hkby0zhenfu}%
\end{align}
The heat kernel is now expressed by the forward scattering amplitude.

Similarly, by Eq.(\ref{wsandrehe}) we can express the one-loop effective
action by the forward scattering amplitude,%
\begin{align}
W\left(  s\right)   &  =-\frac{1}{2}\tilde{\mu}^{2s}\int_{0}^{\infty
}\mathrm{d}tt^{s-1}\left[  \frac{2}{\pi}t\int_{0}^{\infty}f\left(  0\right)
\mathrm{e}^{-k^{2}t}k^{2}\mathrm{d}k\right] \nonumber\\
&  =-\tilde{\mu}^{2s}\frac{\Gamma\left(  s+1\right)  }{\pi}\int_{0}^{\infty
}f\left(  0\right)  \left(  k^{2}\right)  ^{-s}\mathrm{d}k.
\end{align}
By Eq.(\ref{eandrehe}), we can express the vacuum energy by the forward
scattering amplitude,%
\begin{align}
E_{0}\left(  \epsilon\right)   &  =\frac{1}{2}\tilde{\mu}^{2\epsilon}\frac
{1}{\Gamma\left(  -\frac{1}{2}+\epsilon\right)  }\int_{0}^{\infty}%
\mathrm{d}tt^{-\frac{1}{2}+\epsilon-1}\left[  \frac{2}{\pi}t\int_{0}^{\infty
}f\left(  0\right)  \mathrm{e}^{-k^{2}t}k^{2}\mathrm{d}k\right] \nonumber\\
&  =\tilde{\mu}^{2\epsilon}\frac{1}{\pi}\frac{\Gamma\left(  \epsilon+\frac
{1}{2}\right)  }{\Gamma\left(  \epsilon-\frac{1}{2}\right)  }\int_{0}^{\infty
}f\left(  0\right)  \left(  k^{2}\right)  ^{\frac{1}{2}-\epsilon}\mathrm{d}k.
\end{align}

We can also provide the spectral zeta function expressed in terms of the
scattering amplitude. By the relation between $\delta\left(  q\right)  $ and
the spectral zeta function \cite{beauregard2015casimir},%
\begin{equation}
\zeta\left(  s\right)  =\frac{2s}{\pi}\int_{0}^{\infty}q\left(  q^{2}%
+m^{2}\right)  ^{-s-1}\delta\left(  q\right)  q\mathrm{d}q
\end{equation}
and%
\begin{equation}
\delta\left(  q\right)  =-\frac{\pi}{2\pi\mathrm{i}}\int_{C-\mathrm{i}\infty
}^{C+\mathrm{i}\infty}q^{-s}\frac{\zeta_{0}\left(  -\frac{s}{2}\right)  }%
{s}\mathrm{d}s,
\end{equation}
we can connect the spectral zeta function to the scattering amplitude:%
\begin{equation}
\zeta\left(  s\right)  =\frac{2s}{\pi}\int_{0}^{\infty}k\left(  k^{2}%
+m^{2}\right)  ^{-s-1}f\left(  0\right)  k^{2}\mathrm{d}k.
\end{equation}

As a verification, substituting the first-order Born approximation for the
scattering amplitude \cite{sakurai1995modern},%
\begin{equation}
f_{\text{Born}}^{\left(  1\right)  }\left(  \theta\right)  =-\frac{1}{q}%
\int_{0}^{\infty}\mathrm{d}rrV\left(  r\right)  \sin\left(  qr\right)  ,
\end{equation}
where $q=2k\sin\frac{\theta}{2}$, into Eq. (\ref{hkby0zhenfu}) gives%
\begin{equation}
K\left(  t\right)  =\frac{2}{\pi}t\int_{0}^{\infty}\left[  -\frac{1}{q}%
\int_{0}^{\infty}\mathrm{d}rrV\left(  r\right)  \sin\left(  qr\right)
\right]  _{\theta=0}\mathrm{e}^{-k^{2}t}k^{2}\mathrm{d}k. \label{ktandf}%
\end{equation}
For $\theta=0$, $\frac{\sin\left(  qr\right)  }{q}=r$. Substituting into Eq.
(\ref{ktandf}) give%
\begin{align}
K\left(  t\right)   &  =-\frac{2}{\pi}t\int_{0}^{\infty}\int_{0}^{\infty
}\mathrm{d}rr^{2}V\left(  r\right)  \mathrm{e}^{-k^{2}t}k^{2}\mathrm{d}%
k\nonumber\\
&  =-\frac{1}{\sqrt{4\pi t}}\int_{0}^{\infty}V\left(  r\right)  r^{2}%
\mathrm{d}r.
\end{align}
This agrees with the result given by Eq. (\ref{ktborn1}).

Moreover, similar calculations give the one-loop effective action,%
\begin{equation}
W\left(  s\right)  =\frac{\tilde{\mu}^{2s}\Gamma\left(  s-\frac{1}{2}\right)
}{4\sqrt{\pi}}\left(  m^{2}\right)  ^{\frac{1}{2}-s}\int_{0}^{\infty}V\left(
r\right)  r^{2}\mathrm{d}r
\end{equation}
and the vacuum energy%
\begin{equation}
E_{0}\left(  \epsilon\right)  =-\frac{\tilde{\mu}^{2\epsilon}\Gamma
(\epsilon-1)}{4\sqrt{\pi}\Gamma\left(  \epsilon-\frac{1}{2}\right)  }\left(
m^{2}\right)  ^{1-\epsilon}\int_{0}^{\infty}V\left(  r\right)  r^{2}%
\mathrm{d}r.
\end{equation}

\section{Conclusion \label{Conclusion}}

One-loop effective actions and vacuum energies in quantum field theory,
scattering phase shifts and scattering amplitudes in quantum mechanics,
partition functions, and various thermodynamic quantities in statistical
mechanics are all spectral functions. By identifying the relationship between
these spectral functions, it becomes possible to translate the methodology of
calculating one spectral function into the methodology of calculating another.
In this paper, we demonstrate the conversion of the scattering method in
quantum mechanics into the corresponding method in quantum field theory.
Specifically, we convert the Born approximation and the WKB approximation into
methodologies for calculating the one-loop effective action and vacuum energy.
Theoretically, all methodologies for calculating scattering phase shifts and
amplitude in quantum mechanics, such as the eikonal approximation, can be
converted into methodologies for calculating effective actions and vacuum energies.

This approach can calculate various spectral functions across different
physical domains. Spectral functions in quantum field theory, quantum
mechanics, and statistical mechanics can be transformed into one another using
their corresponding relations. As such, methods utilized in one physical
domain can be converted into methods in another. For instance, determining the
energy spectrum of an interacting many-body system is a fundamental problem in
statistical mechanics. Eigenvalues are the most basic spectral functions and
can be calculated from other spectral functions. In Ref.
\cite{zhou2018calculating}, for example, the energy eigenvalue of an
interacting many-body system is calculated using the partition function.
Various statistical mechanics methods have been developed to calculate
partition and grand partition functions, such as the cluster expansion method,
field theory method \cite{pathria2011statistical}, and some mathematical
methods \cite{zhou2018canonical,zhou2018statistical}. Both the eigenvalue and
partition function are spectral functions. Methods for calculating partition
functions, such as the cluster expansion method, can be transformed into
methods for calculating the energy spectrum of interacting gases
\cite{li2022energy}.

In quantum-mechanical scattering theory, the focus is primarily on short-range
scattering, although the Born approximation can handle certain long-range
potentials, such as the Coulomb potential. For long-range potential
scattering, the scattering phase shift can be uniformly treated using the
tortoise coordinate \cite{li2021long}. Future works will delve into this approach.

A duality exists in classical and quantum mechanics, as well as in field
theory
\cite{li2021duality,liu2021exactly,chen2022indirect,li2022solving,gu2025exactly,gu2024exactly,gu2022duality}%
. In quantum mechanics, this duality pertains to the relationship between
various eigenproblems, whereas in field theory, it pertains to the
relationship between different fields. In this paper, we establish a link
between spectral functions in quantum mechanics and quantum field theory.
Future works will delve further into the relationship between these problems.

In summary, this paper proposes a methodology for transforming one spectral
function problem into another spectral function problem through the use of
spectral function transformations. This approach enables the conversion of
methods from different areas of physics, such as quantum field theory, quantum
mechanics, and statistical mechanics, into one another.

\section{Integral representation of Bessel function
\label{integralrepresentation}}

In this appendix, we give some integral representations for the Bessel function.

\subsection{Integral representation of $j_{l}^{2}\left(  kr\right)  $
\label{Appendix1}}

Taking $\left\vert \mathbf{u}\right\vert =\left\vert \mathbf{v}\right\vert
=kr$ in the expansion \cite{watson1944treatise}
\begin{equation}
\frac{\sin w}{w}=\sum_{l=0}^{\infty}\left(  2l+1\right)  j_{l}\left(
v\right)  j_{l}\left(  u\right)  P_{l}\left(  \cos\theta\right)  ,
\label{besselj}%
\end{equation}
where $w=\sqrt{u^{2}+v^{2}-2uv\cos\theta}$ and $\theta$ is the angle between
$\mathbf{u}$ and $\mathbf{v}$, gives
\begin{equation}
\frac{\sin qr}{qr}=\sum_{l=0}^{\infty}\left(  2l+1\right)  j_{l}^{2}\left(
kr\right)  P_{l}\left(  \cos\theta\right)  \label{sinqrqr}%
\end{equation}
with $w=qr=2kr\sin\frac{\theta}{2}$. Multiplying $P_{l^{\prime}}\left(
\cos\theta\right)  $ on both sides of Eq. (\ref{sinqrqr}) and then integrating
from $0$ to $\pi$ give
\begin{equation}
\int_{0}^{\pi}\frac{\sin qr}{qr}P_{l^{\prime}}\left(  \cos\theta\right)
\sin\theta\mathrm{d}\theta=\int_{0}^{\pi}\sum_{l=0}^{\infty}\left(
2l+1\right)  j_{l}^{2}\left(  kr\right)  P_{l}\left(  \cos\theta\right)
P_{l^{\prime}}\left(  \cos\theta\right)  \sin\theta\mathrm{d}\theta.
\end{equation}
By%
\begin{equation}
\int_{0}^{\pi}P_{l}\left(  \cos\theta\right)  P_{l^{\prime}}\left(  \cos
\theta\right)  \sin\theta\mathrm{d}\theta=\frac{2}{2l+1}\delta_{ll^{\prime}},
\end{equation}
we have
\begin{equation}
\int_{0}^{\pi}\frac{\sin qr}{qr}P_{l^{\prime}}\left(  \cos\theta\right)
\sin\theta\mathrm{d}\theta=2j_{l^{\prime}}^{2}\left(  kr\right)  .
\end{equation}
This gives an integral representation of $j_{l}^{2}\left(  kr\right)  $:
\begin{equation}
j_{l}^{2}\left(  kr\right)  =\frac{1}{2}\int_{0}^{\pi}\frac{\sin qr}{qr}%
P_{l}\left(  \cos\theta\right)  \sin\theta\mathrm{d}\theta,
\end{equation}
where $l$ is an integer.

\subsection{Integral representation of $j_{l}\left(  kr\right)  n_{l}\left(
kr\right)  $ \label{Appendix2}}

Taking $\left\vert \mathbf{u}\right\vert =\left\vert \mathbf{v}\right\vert
=kr$ in the expansion \cite{watson1944treatise}%
\begin{equation}
\frac{\cos w}{w}=-\sum_{l=0}^{\infty}\left(  2l+1\right)  j_{l}\left(
v\right)  n_{l}\left(  u\right)  P_{l}\left(  \cos\theta\right)  ,
\end{equation}
where $w=\sqrt{u^{2}+v^{2}-2uv\cos\theta}$ and $\theta$ is the angle between
$\mathbf{u}$ and $\mathbf{v}$, gives%
\begin{equation}
\frac{\cos qr}{qr}=-\sum_{l=0}^{\infty}\left(  2l+1\right)  j_{l}\left(
kr\right)  n_{l}\left(  kr\right)  P_{l}\left(  \cos\theta\right)
\label{besselj n1}%
\end{equation}
with $w=qr=2kr\sin\frac{\theta}{2}$. Multiplying $P_{l^{\prime}}\left(
\cos\theta\right)  $ on both sides of Eq. (\ref{besselj n1}) and then
integrating from $0$ to $\pi$ give
\begin{align}
\int_{0}^{\pi}\frac{\cos qr}{qr}P_{l^{\prime}}\left(  \cos\theta\right)
\sin\theta\mathrm{d}\theta &  =-\int_{0}^{\pi}\sum_{l=0}^{\infty}\left(
2l+1\right)  j_{l}\left(  kr\right)  n_{l}\left(  kr\right)  P_{l}\left(
\cos\theta\right)  P_{l^{\prime}}\left(  \cos\theta\right)  \sin
\theta\mathrm{d}\theta\nonumber\\
&  =-2j_{l^{\prime}}\left(  kr\right)  n_{l^{\prime}}\left(  kr\right)  .
\end{align}
This gives an integral representation of $j_{l}\left(  kr\right)  n_{l}\left(
kr\right)  $:
\begin{equation}
j_{l}\left(  kr\right)  n_{l}\left(  kr\right)  =-\frac{1}{2}\int_{0}^{\pi
}\frac{\cos qr}{qr}P_{l}\left(  \cos\theta\right)  \sin\theta\mathrm{d}\theta,
\end{equation}
where $l$ is an integer.

\subsection{Integral representation of $J_{l+\mu}^{2}\left(  kr\right)  $
\label{Appendix3}}

Taking $\left\vert \mathbf{u}\right\vert =\left\vert \mathbf{v}\right\vert
=kr$ in the expansion \cite{watson1944treatise}%
\begin{equation}
\frac{J_{\mu}\left(  w\right)  }{w^{\mu}}=2^{\mu}\Gamma\left(  \mu\right)
\sum_{l=0}^{\infty}\left(  l+\mu\right)  \frac{J_{l+\mu}\left(  u\right)
J_{l+\mu}\left(  v\right)  }{u^{\mu}v^{\mu}}C_{l}^{\mu}\left(  \cos
\theta\right)  ,\text{ \ \ }\left(  u>v\right)  , \label{besseljj}%
\end{equation}
where $w=\sqrt{u^{2}+v^{2}-2uv\cos\theta}=qr=2kr\sin\frac{\theta}{2}$ and
$C_{l}^{\mu}\left(  z\right)  $ is the Gegenbauer polynomial
\cite{olver2010nist}, multiplying $C_{l^{\prime}}^{\mu}\left(  \cos
\theta\right)  $ on both sides of\ Eq. (\ref{besseljj}), and integrating from
$0$ to $\pi$ give%
\begin{equation}
\int_{0}^{\pi}\frac{J_{\mu}\left(  qr\right)  }{q^{\mu}r^{\mu}}C_{l^{\prime}%
}^{\mu}\left(  \cos\theta\right)  \sin^{2\mu}\theta\mathrm{d}\theta=2^{\mu
}\Gamma\left(  \mu\right)  \sum_{l=0}^{\infty}\left(  l+\mu\right)
\frac{J_{l+\mu}^{2}\left(  kr\right)  }{\left(  kr\right)  ^{2\mu}}\int%
_{0}^{\pi}C_{l}^{\mu}\left(  \cos\theta\right)  C_{l^{\prime}}^{\mu}\left(
\cos\theta\right)  \sin^{2\mu}\theta\mathrm{d}\theta.
\end{equation}
By \cite{olver2010nist}
\begin{equation}
\int_{0}^{\pi}C_{l}^{\mu}\left(  \cos\theta\right)  C_{l^{\prime}}^{\mu
}\left(  \cos\theta\right)  \sin^{2\mu}\theta\mathrm{d}\theta=\frac{2^{1-2\mu
}\pi\Gamma\left(  l+2\mu\right)  }{\Gamma\left(  l+1\right)  \left(
l+\mu\right)  \Gamma^{2}\left(  \mu\right)  }\delta_{ll^{\prime}},
\end{equation}
we arrive at an integral representation of $J_{l+\mu}^{2}\left(  kr\right)
$:
\begin{equation}
J_{l+\mu}^{2}\left(  kr\right)  =\frac{2^{\mu-1}\Gamma\left(  l+1\right)
\Gamma\left(  \mu\right)  }{\pi\Gamma\left(  2\mu+l\right)  }\left(
kr\right)  ^{2\mu}\int_{0}^{\pi}\frac{J_{\mu}\left(  qr\right)  }{q^{\mu
}r^{\mu}}C_{l}^{\mu}\left(  \cos\theta\right)  \sin^{2\mu}\theta
\mathrm{d}\theta,
\end{equation}
where $l$ is an integer and $\mu$ is a real number.

\subsection{Integral representation of $J_{l+\nu}\left(  kr\right)  Y_{l+\nu
}\left(  kr\right)  $ \label{Appendix4}}

Taking $\left\vert \mathbf{u}\right\vert =\left\vert \mathbf{v}\right\vert
=kr$ in the expansion \cite{watson1944treatise}%
\begin{subequations}
\begin{equation}
\frac{Y_{\mu}\left(  w\right)  }{w^{\mu}}=2^{\mu}\Gamma\left(  \mu\right)
\sum_{l=0}^{\infty}\left(  l+\mu\right)  \frac{Y_{l+\mu}\left(  u\right)
J_{l+\mu}\left(  v\right)  }{u^{\mu}v^{\mu}}C_{l}^{\mu}\left(  \cos
\theta\right)  ,\text{ \ \ }\left(  u>v\right)  , \label{besseljn2}%
\end{equation}
where $w=\sqrt{u^{2}+v^{2}-2uv\cos\theta}=$ $qr=2kr\sin\frac{\theta}{2}$ and,
multiplying $C_{l^{\prime}}^{\mu}\left(  \cos\theta\right)  $ on both sides of
Eq. (\ref{besseljn2}) and integrating from $0$ to $\pi$ give%
\end{subequations}
\begin{equation}
\int_{0}^{\pi}\frac{Y_{\mu}\left(  qr\right)  }{q^{\mu}r^{\mu}}C_{l^{\prime}%
}^{\mu}\left(  \cos\theta\right)  \sin^{2\mu}\theta\mathrm{d}\theta=2^{\mu
}\Gamma\left(  \mu\right)  \sum_{l=0}^{\infty}\left(  l+\mu\right)
\frac{J_{l+\mu}\left(  kr\right)  Y_{l+\mu}\left(  kr\right)  }{\left(
kr\right)  ^{2\mu}}\int_{0}^{\pi}C_{l}^{\mu}\left(  \cos\theta\right)
C_{l^{\prime}}^{\mu}\left(  \cos\theta\right)  \sin^{2\mu}\theta
\mathrm{d}\theta.
\end{equation}
By \cite{olver2010nist}
\begin{equation}
\int_{0}^{\pi}C_{l}^{\mu}\left(  \cos\theta\right)  C_{l^{\prime}}^{\mu
}\left(  \cos\theta\right)  \sin^{2\mu}\theta\mathrm{d}\theta=\frac{2^{1-2\mu
}\pi\Gamma\left(  l+2\mu\right)  }{\Gamma\left(  l+1\right)  \left(
l+\mu\right)  \Gamma^{2}\left(  \mu\right)  }\delta_{ll^{\prime}},
\end{equation}
we arrive at an integral representation of $J_{l+\mu}\left(  kr\right)
Y_{l+\mu}\left(  kr\right)  $:
\begin{equation}
J_{l+\mu}\left(  kr\right)  Y_{l+\mu}\left(  kr\right)  =\frac{2^{\mu-1}%
\Gamma\left(  l+1\right)  \Gamma\left(  \mu\right)  }{\pi\Gamma\left(
2\mu+l\right)  }\left(  kr\right)  ^{2\mu}\int_{0}^{\pi}\frac{Y_{\mu}\left(
qr\right)  }{q^{\mu}r^{\mu}}C_{l}^{\mu}\left(  \cos\theta\right)  \sin^{2\mu
}\theta\mathrm{d}\theta,
\end{equation}
where $l$ is an integer and $\mu$ is a real number.

\section*{Acknowledgements}
We are very indebted to Dr G. Zeitrauman for his encouragement. This work is supported
in part by Special Funds for Theoretical Physics Research Program of the NSFC under
Grant No. 11947124, and NSFC under Grant Nos. 11575125, and 11675119.











\end{document}